\documentclass[preprint,12pt]{elsarticle}

\usepackage{graphicx}

\usepackage{amssymb}

\usepackage{lineno}
\journal{Nuclear Instruments and Methods in Physics Research A}

\begin{document}

\begin{frontmatter}



\title{Multiplicative scale uncertainties in the unified approach for constructing confidence intervals}


\author{E.S. Smith\corref{corauthor}}
\address{Thomas Jefferson National Accelerator Facility, Newport News, Virginia 23606}
\cortext[corauthor]{Corresponding author. E-mail address: elton@jlab.org.}


\begin{abstract}
We have investigated how uncertainties in the estimation of the detection 
efficiency affect the 90\% confidence intervals in the unified approach 
for constructing confidence intervals. The study has been conducted for
experiments where the number of detected events is
large and can be described by a Gaussian probability density function.
We also assume the detection efficiency has a Gaussian probability density and study
the range of the relative uncertainties  $\sigma_\epsilon$ between 0 and 30\%.
We find that the confidence intervals provide proper coverage over a wide signal range and increase 
smoothly and continuously from the intervals that ignore scale uncertainties 
with a quadratic dependence on $\sigma_\epsilon$.

\end{abstract}


\end{frontmatter}


\section{Introduction}
In the search for new physics, experiments select a region of interest, the ``signal region,'' where 
new phenomena are expected. The search consists of selecting a number of 
events with characteristics of the expected signal, but the selection 
process usually has a finite detection efficiency and includes events  from uninteresting
known sources, generically  called background. In the absence
of a significant excess in the detected number of events over the expected
background, we use the data to set limits on the range of the possible 
signal at a certain confidence level. 
The confidence intervals depend not only on the number of detected events, but also
on the precision with which one knows the 
detection efficiency and the average background rate in the signal region. Therefore, the uncertainties in these quantities will
affect the sensitivity with which limits can be set by a particular measurement.
The efficiency and background level are examples of what are referred to in the literature
as {\em nuisance}  parameters, and the precision with which they are known are measures 
of systematic uncertainties that degrade the sensitivity with which the experiment can set limits on
the signal of interest.  For the 
present study, we do not address how the efficiency or backgrounds are obtained, but we
assume that estimates for these quantities and their uncertainties have been determined for
known experimental conditions.

The procedures for including systematic uncertainties in the determination of 
confidence intervals is a topic of intense interest and discussion. For a recent review of
the treatment of systematic errors, see Ref.\,\cite{Heinrich:2007zz}, and for a review of
the treatment of nuisance parameters specifically, see Ref.\,\cite{Cousins:2005hc} and references therein. 
There are two primary methods that have been used to incorporate systematic errors
into the calculation of upper limits. The first makes use of the profile likelihood approach
(see for example Ref.\,\cite{Rolke:2004mj}) in combination with the ``$\ln{\mathfrak{L}}+\frac{1}{2}$ method,'' which
extracts the confidence interval by finding the points where
the $-2$~log likelihood function increases by 2.706 (90\% confidence level).
The profile likelihood is evaluated for values of the parameters of interest that maximize
the likelihood for a specific hypothesis for the signal rate and over the entire range of all other nuisance parameters
of relevance.  The second method, which
we adopt in this study, conducts a full construction
of the confidence interval after folding the primary probability distribution
function with a distribution that describes the nuisance parameters. 
A consensus on which methods are optimal
in specific experimental situations has not been reached, although the field has matured
considerably over the past decade. 

This paper focuses on experiments where the number of detected events is
large and can be described using a Gaussian probability density function.
To date, the studies for inclusion of systematic uncertainties in the determination of 
confidence regions has concentrated  on experiments that record a  
small number of events and whose probability density functions are described by Poisson distributions
\cite{Conrad:2002kn,Tegenfeldt:2004dk}.
The case where a large number of events are detected
in the signal region, consistent with expectations for background, is a simpler
problem but has received reduced attention. In this case, the distribution for the {\em true} number
of excess events above background $\mu$ is Gaussian but physically bounded to be positive.
Of course, the number of {\em measured} events relative to background $x$ can fluctuate above and
below the expected background level and can be both positive and negative.
In principle, this situation is a straightforward application of the case for Poisson statistics.
However, the blind use of those procedures can lead to intense computation and is often
compounded by issues of numerical precision. 

We construct confidence regions following the approach developed by Cousins and collaborators 
and refined by many others. The method follows the 
frequentist method of constructing confidence intervals  proposed by Feldman and Cousins
in Ref.\,\cite{Feldman:1997qc}, but it is modified to include systematic errors following
the original suggestion by Highland and Cousins in Ref. \cite{Cousins:1991qz}.
There are two great advantages of using this unified method
for construction of confidence regions.  First, it avoids unphysical regions
which must be included to interpret classical confidence belts.
Second, it has a smooth transition between a central confidence
belt and upper limits. This has the advantage that it defines a
priori the choice of the confidence interval, independent of measured
data.  In the absence of systematic errors, the 
Feldman-Cousins prescription coincides with the classical Gaussian central
confidence belt for positive measurements far from zero. Below that
there is a smooth transition to single-sided upper limits, corresponding to
confidence intervals that are non-zero for all
values of the measurement~$x$. 

This Neyman construction of confidence intervals is classical or frequentist,  but
the systematic uncertainties are included via a Bayesian procedure.
Systematics are incorporated into the construction by
weighting the probability  distribution with an assumed density for the nuisance parameter
(in our case, the background level and the detection efficiency) and then integrating over the nuisance parameter.
We assume the nuisance parameters have Gaussian probability
densities. Integration over the nuisance parameters is Bayesian as it 
incorporates information about our belief about the detection efficiency and the background
quantities. 

\section{Probability distribution \label{sec:prob}}  
For an experiment with low statistics, the confidence limits
are a function of both the number of observed events (signal plus background) and the level of expected
background. The correct statistical behavior depends on both quantities, since
the difference of two Poisson distributions is no longer a Poisson distribution. In 
the limit of Gaussian statistics, the confidence region is determined only by the excess of events
over background relative to its error, which is dimensionless. This is because the distribution of the measured
signal (difference between the number of observed events and the estimated background) 
is also a Gaussian distribution with the width given by the quadrature sum of
the error of the observed events and the error in the estimated background level. 
The probability distribution for the number of observed events $X$ with statistical uncertainty $\sigma_X$ 
representing a fixed but unknown signal $\mu$ and a background level $b$ is given by
\begin{eqnarray}
P(X,b,\mu,\mu_b) & = & {1 \over \sqrt{2 \pi\sigma_X^2}} \;  {1 \over \sqrt{2 \pi\sigma_b^2}}\;  \nonumber \\
& & \exp\left[-{ {(X-b-\mu)^2} \over {2\sigma_X^2}}\right]
\exp\left[-{ {(b-\mu_b)^2} \over {2\sigma_{\mu_b}^2}}\right],
\end{eqnarray}
where $\mu_b$ is the estimated value for $b$ and has an uncertainty  $\sigma_{\mu_b}$.
The probability density for the number of signal events $x=X - b$, or excess above background, is then obtained
by integrating over the assumed distribution of the true background level $b$:
\begin{eqnarray}
P(x,\mu) & = & \int_{-\infty}^{\infty} P(X,b,\mu,\mu_b)  \; db \\  
 & = & {1 \over \sqrt{2 \pi\sigma^2}} \; \exp\left[-{ {(x-\mu)^2} \over {2\sigma^2}}\right], \label{eq:gprob} \end{eqnarray}
where $\sigma  =  \sqrt{\sigma_X^2 + \sigma_{\mu_b}^2}$ is the statistical uncertainty associated with the extraction
of the signal.

\begin{figure}[tp]
\begin{center}
\includegraphics[height=7cm,bb=20 0 500 520,clip=true]{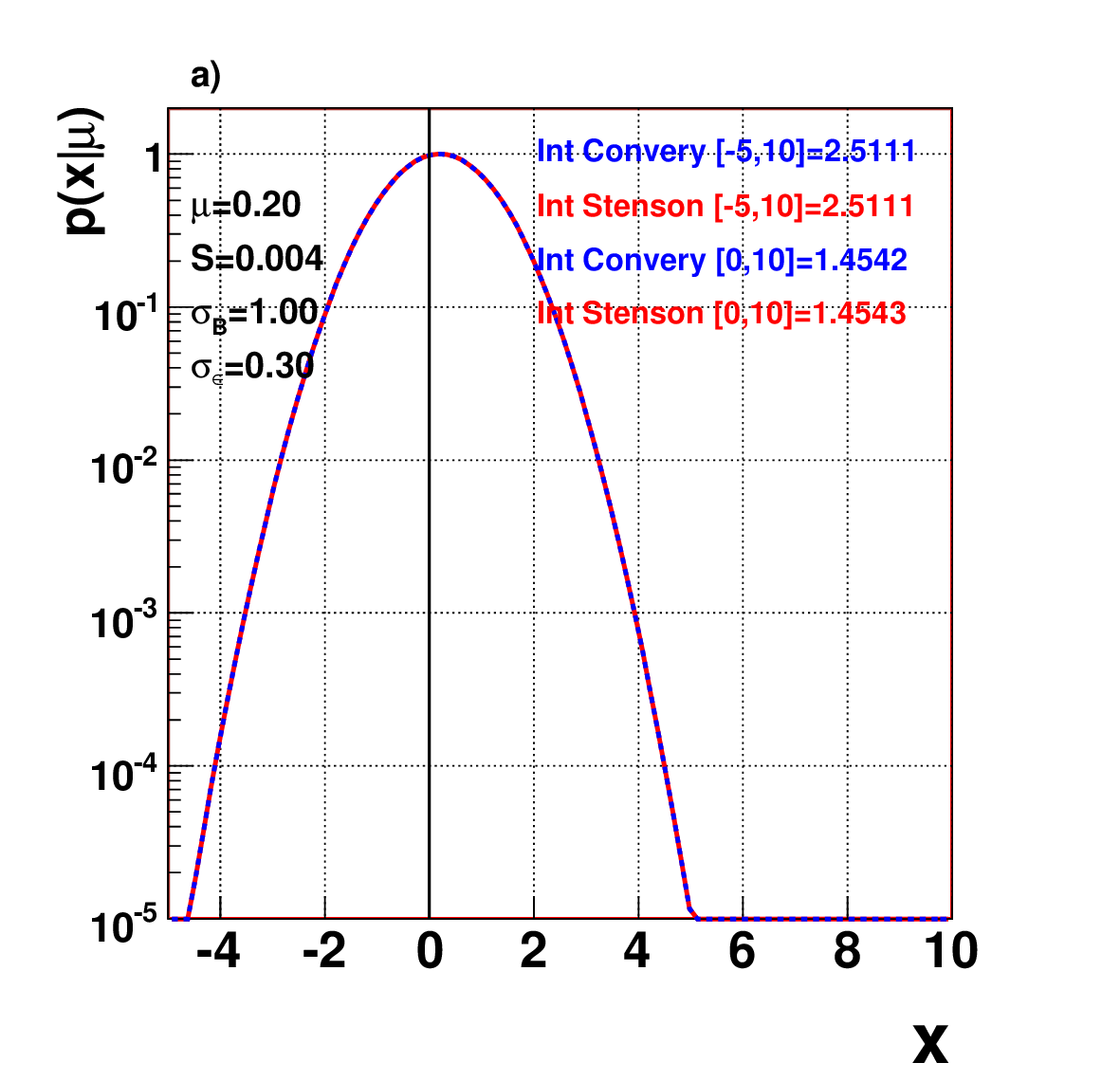}
\includegraphics[height=7cm,bb=20 0 500 520,clip=true]{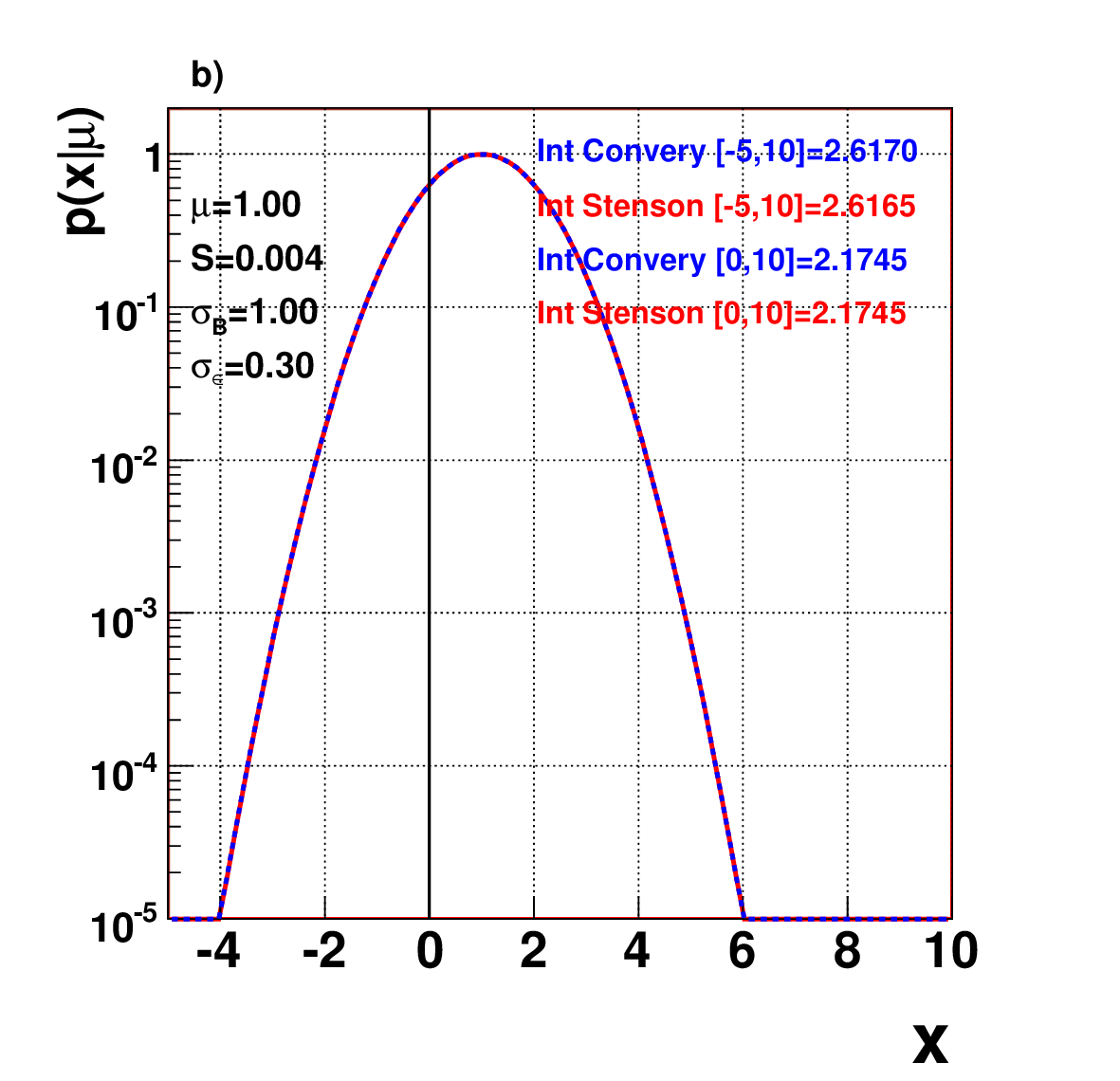} \\*[0.5cm]
\includegraphics[height=7cm,bb=20 0 500 520,clip=true]{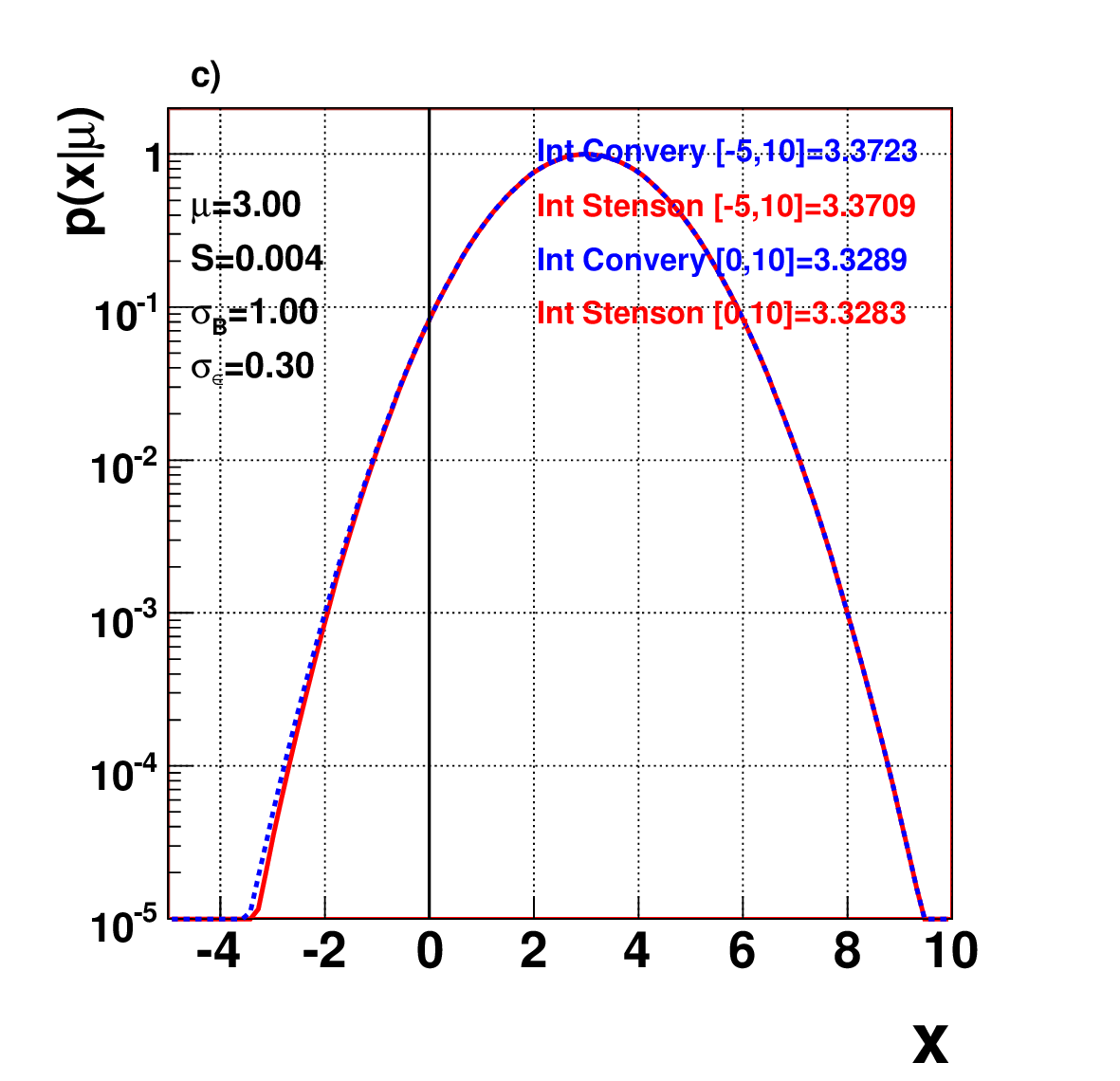}
\includegraphics[height=7cm,bb=20 0 500 520,clip=true]{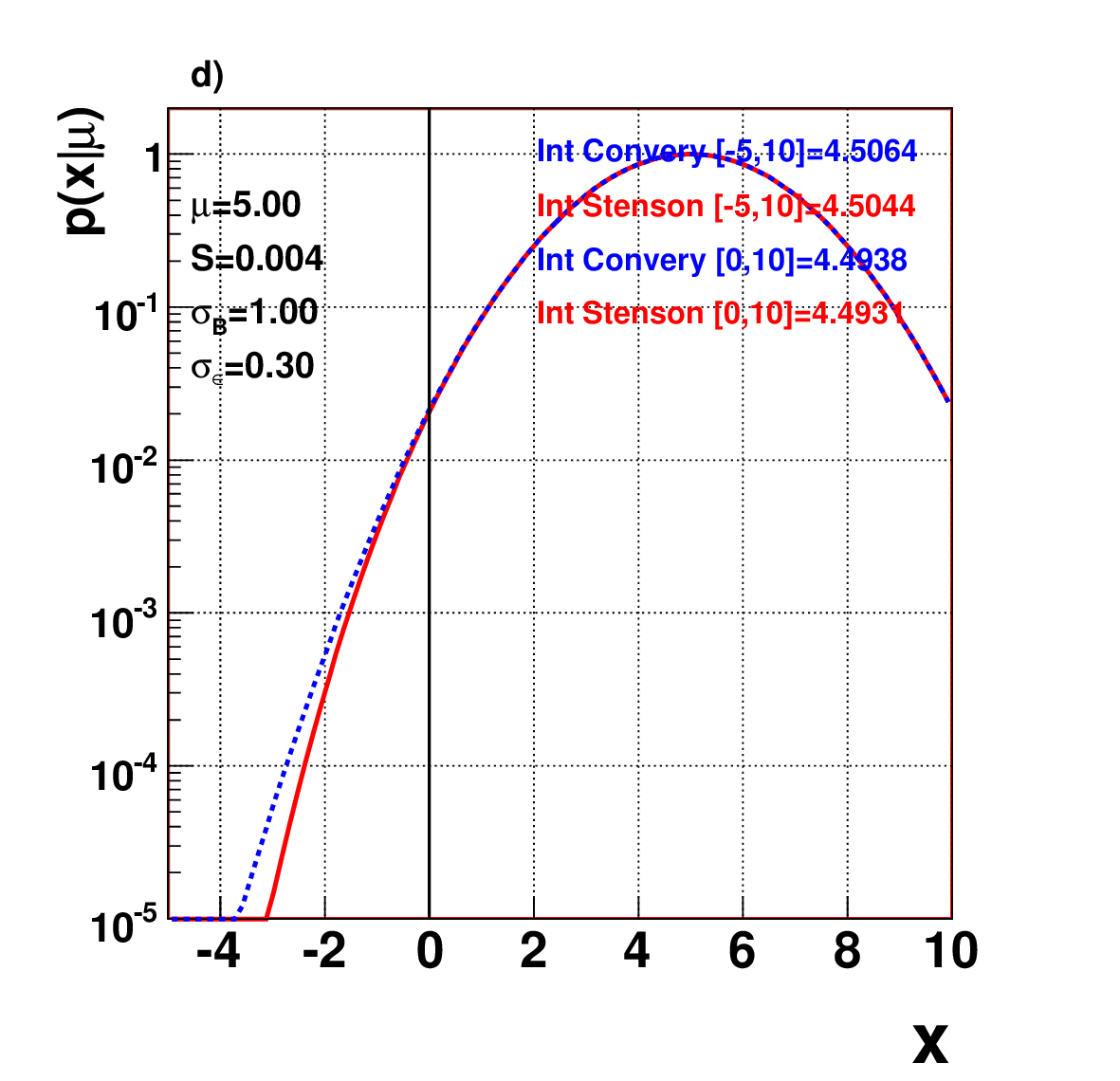}
\caption{(Color online) The probability distribution is plotted for $\sigma_\epsilon$=0.3 and $\hat{S}$=0.004 and four different values of the 
true mean $\mu$. The exact solution by Stenson \cite{stenson} is shown (solid line) along with
the approximate solution by Convery \cite{convery} (dashed line). The function is approximately Gaussian with a modified
width for all values of the parameters. The curves practically coincide in the top plots, and  the integrals of the curves are 
compared for two different intervals, showing that the approximate solution is very good.
\label{fig:plot_multsys3}}
\end{center}     
\end{figure}

If we include the uncertainty $\sigma_S$ in the efficiency $S$ for detecting the signal, the distribution
is given by the following:
\begin{eqnarray}
P(x,\mu) & \propto & \int_0^1 \exp\left[-{ {(S\mu/\hat{S}-x)^2} \over {2\sigma^2}}\right]
\exp\left[-{ {(S-\hat{S})^2} \over {2\sigma_{S}^2}}\right]    \; dS, \label{eq:background}
\end{eqnarray}
where $\hat{S}$ is the measured detection efficiency, and the integral is performed over the posteriori Gaussian probability
for the true efficiencies $S$. We note that the integration limits lead to artificial boundaries, which are discussed below.
We will also refer to $\hat{S}$ as the ``scale factor''
for the signal. Defining  the relative uncertainty as $\sigma_\epsilon = \sigma_S/\hat{S} $, 
the probability distribution is given by
\begin{eqnarray}
P(x/\sigma,\mu/\sigma,\sigma_\epsilon,\hat{S}) & \propto & {\sigma\sigma_\epsilon \over \sqrt{2(\mu^2\sigma_\epsilon^2+\sigma^2)} }\; 
\exp\left[- {(\mu-x)^2 \over 2(\mu^2\sigma_\epsilon^2+\sigma^2)}\right] \times \label{eq:fullprob} \\
& & \left\{ {\rm Erf} \left[ {\mu x\,\sigma_\epsilon^2 + \sigma^2} \over {\sigma_\epsilon \sigma \sqrt{2(\mu^2\sigma_\epsilon^2+\sigma^2)}} \right] 
- {\rm Erf} \left[ {(\hat{S}-1)\sigma^2 - \mu \sigma_\epsilon^2 (\mu - x \hat{S}) \over {\hat{S} \sigma_\epsilon \sigma \sqrt{2(\mu^2\sigma_\epsilon^2+\sigma^2)}}} \right] \right\}.
\nonumber
\end{eqnarray}
Convery \cite{convery} derived the expression on the first line of Eq.\,\ref{eq:fullprob}, which is valid for small values of $\sigma_\epsilon$.
The more exact expression requires the expression in curly brackets containing the difference of the two error functions and was obtained by
Stenson \cite{stenson}. We note that the probability distribution converges to Eq.\,\ref{eq:gprob} when $\mu$ is very small.
This is reasonable, since one expects the probability distribution to be unaffected by the scale uncertainty when $\mu$=0.
The probability distributions  for several values of $\mu$  and fixed values of the other parameters are plotted in Fig.\,\ref{fig:plot_multsys3}.
The probability distribution proper must be normalized by setting the integral to unity, and this is done numerically when confidence ranges are calculated.
However, for ease of plotting and comparing the functions for different parameters in the figures, we normalize them by setting $P(\mu)=1$.
From the plots, it is clear that Convery's original expression is a very good approximation over the range of parameters in the figures, as it deviates
slightly from the full distribution by fractions of a percent only for $\sigma_\epsilon$ greater than 20\%.

We make some brief remarks concerning the probability function. The distribution is a 
function of the reduced variables $x/\sigma$ and $\mu/\sigma$, the fractional scale uncertainty
$\sigma_\epsilon$ and the scale itself $\hat{S}$. 
The first \rm{Erf} term affects the tails of the distribution and becomes increasingly important as $\sigma_\epsilon$ increases.
The scale dependence is contained solely in the
second \rm{Erf} term, which evaluates to $-1$ when $\hat{S}$ is small, generally less than 20\%.
The origin of the scale dependence comes from non-physical conditions that occur when the Gaussian distribution for $S$ violate the 
condition $0 \leq S \leq 1$, so they are of little practical interest.
In our examples, we have used $\hat{S} = 0.004$, which is a typical value for the acceptance of some of our experimental searches.
However, the particular value is unimportant, because the probability distribution is independent of $\hat{S}$ in our parameter space, and we
drop its dependence from the argument list.

\section{Ordering principle}
\begin{figure}[tp]
\begin{center}
\includegraphics[height=7cm,bb=20 0 500 520,clip=true]{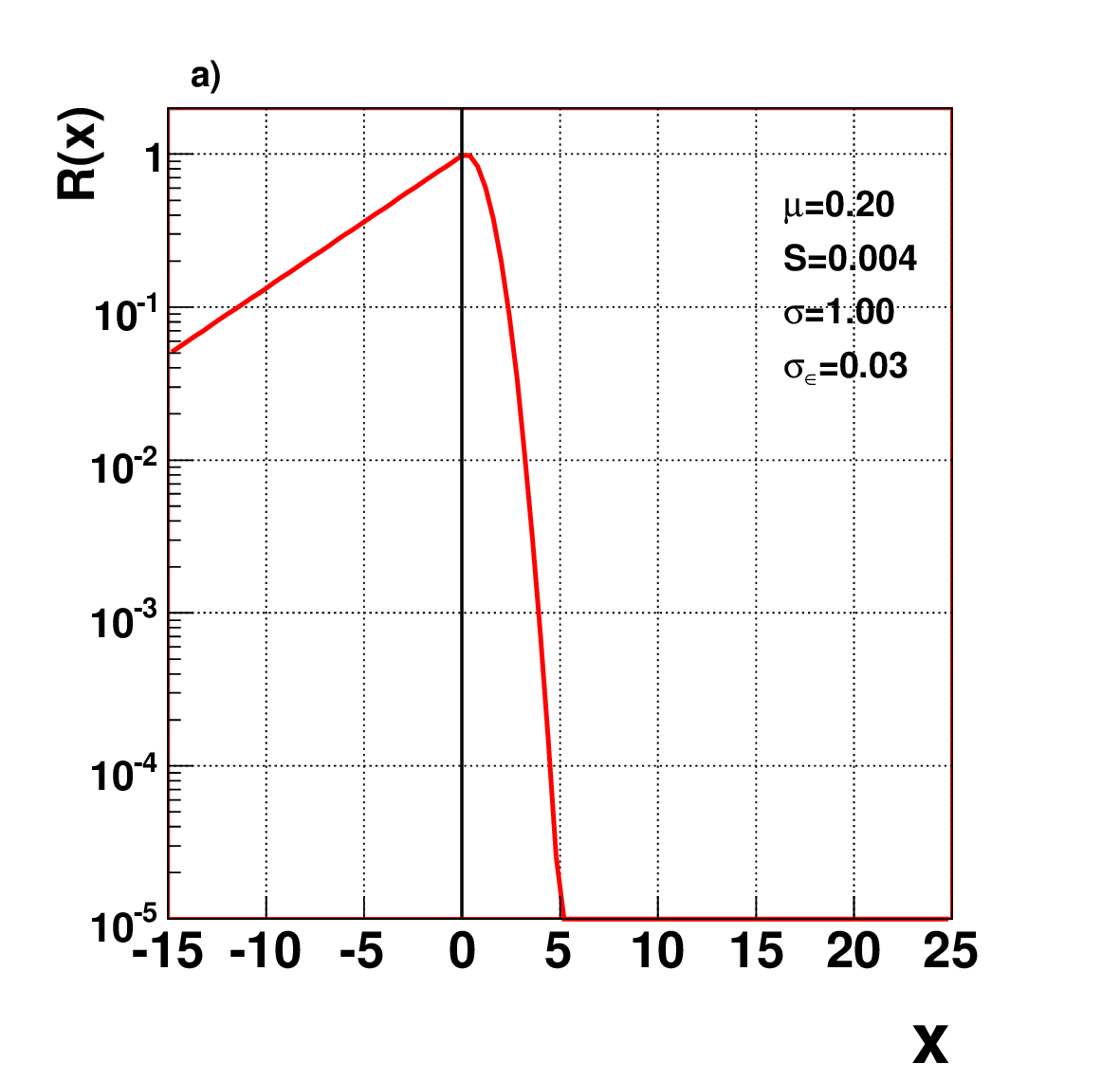}
\includegraphics[height=7cm,bb=20 0 500 520,clip=true]{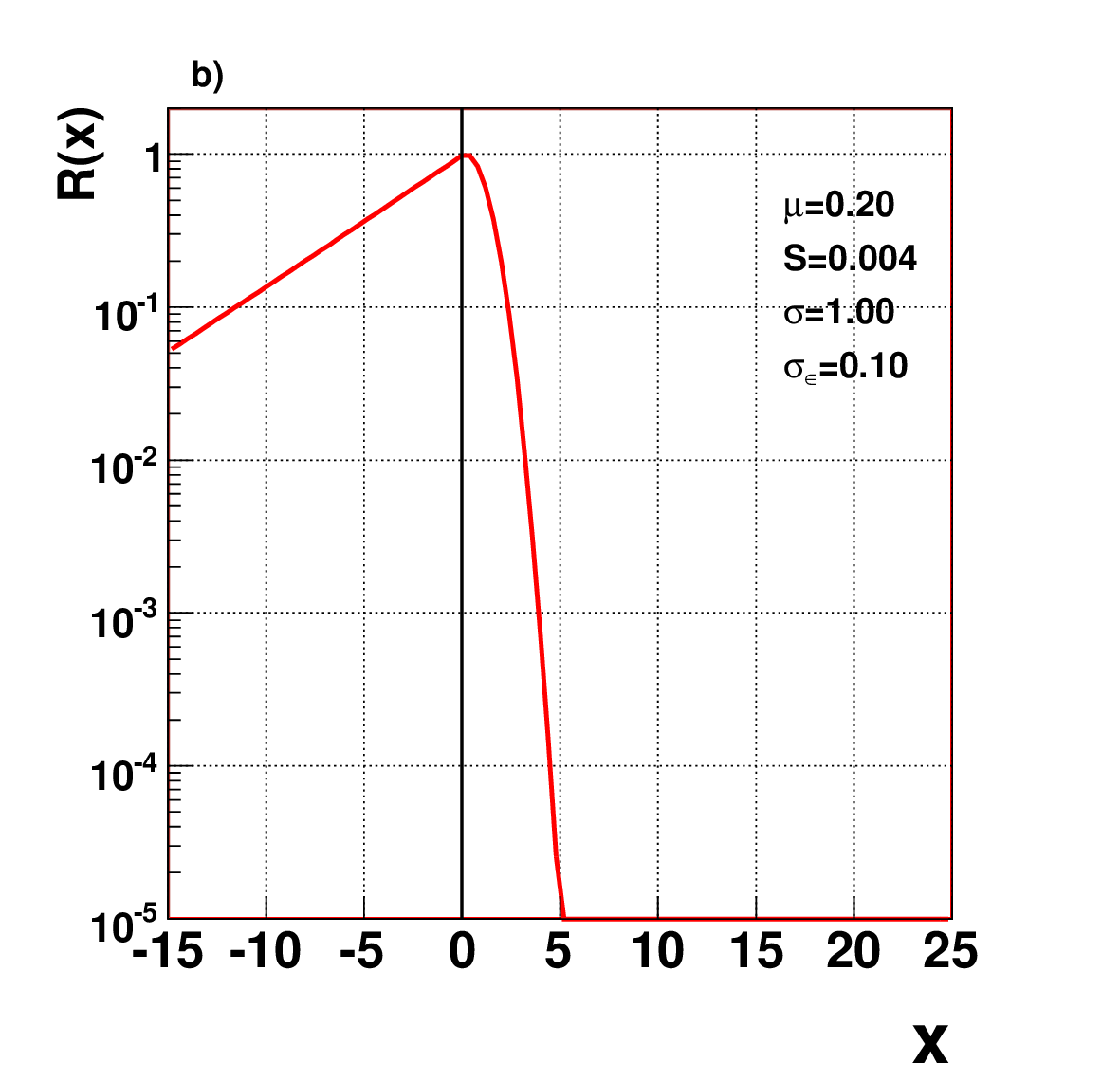} \\*[0.5cm]
\includegraphics[height=7cm,bb=20 0 500 520,clip=true]{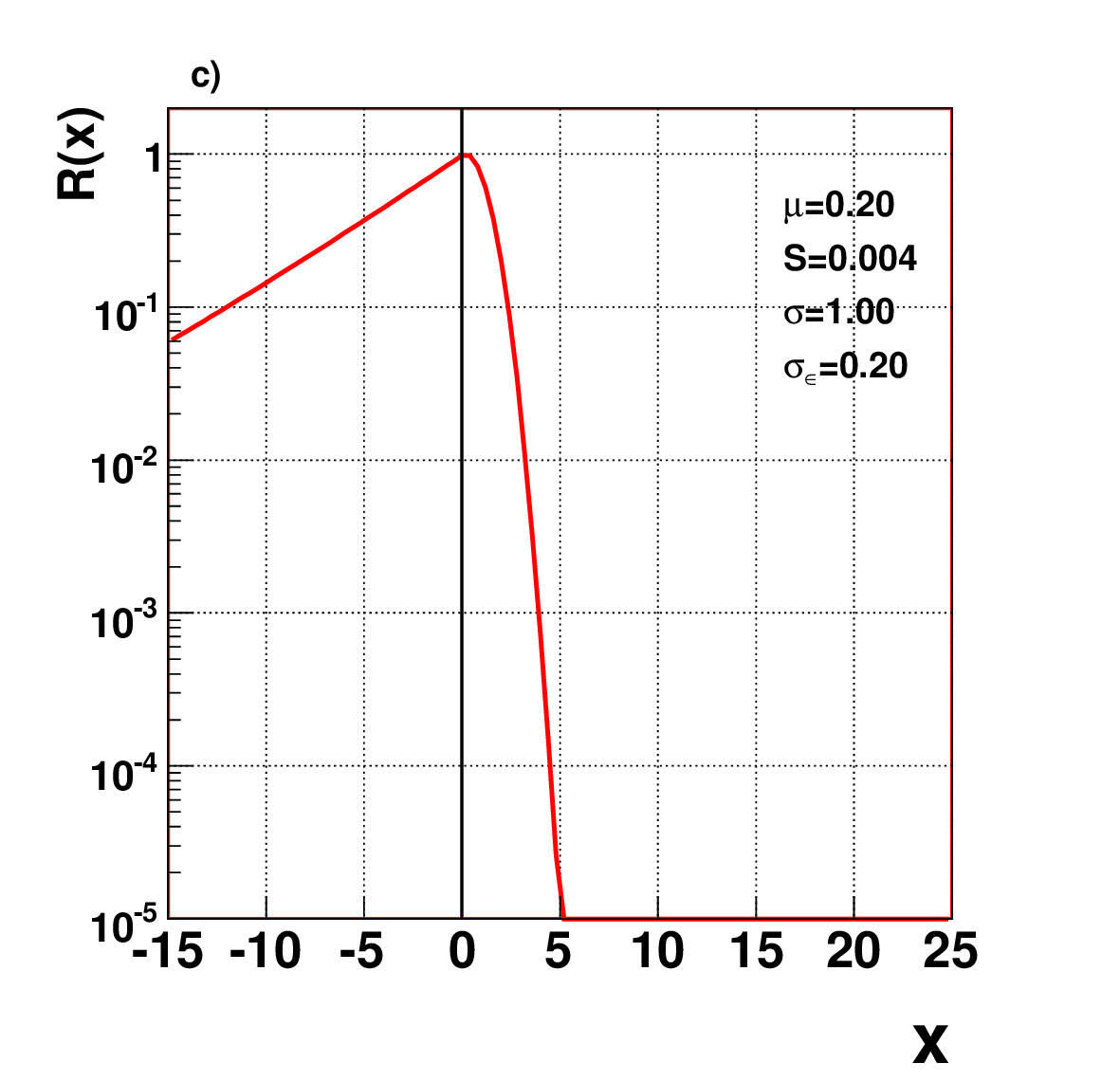}
\includegraphics[height=7cm,bb=20 0 500 520,clip=true]{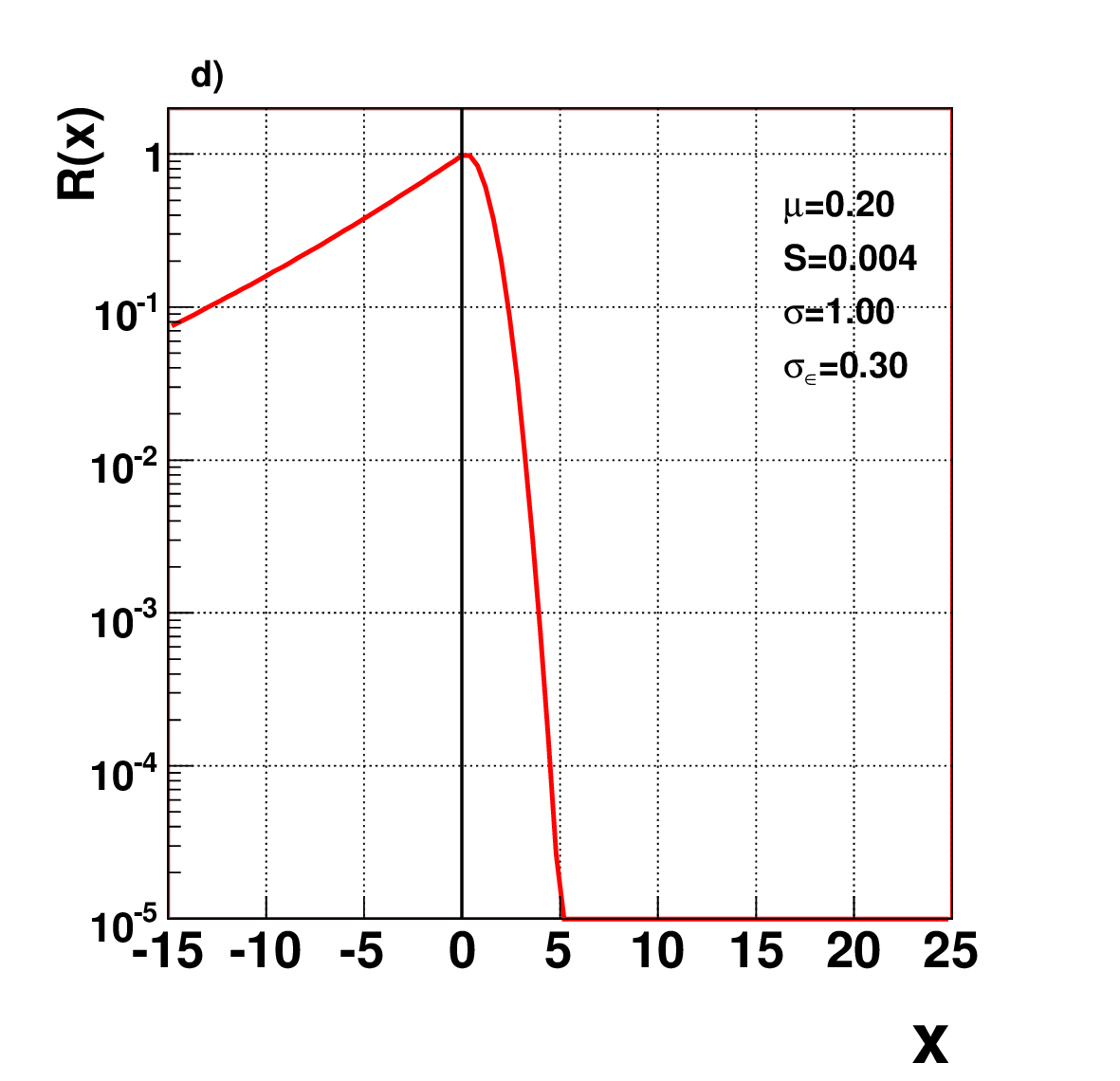}
\caption{(Color online) The ordering principle $R$ is plotted for four different values of the relative scale uncertainty,
$\sigma_\epsilon$, at $\mu$=0.2. The function $R$ develops a tail at low $x$ as
$\sigma_\epsilon$ increases. The tail has the effect of biasing the selection of the confidence interval toward
a one-sided confidence interval. At this low value of $\mu$, the tail is very shallow and extends to very large negative
$x$. 
\label{fig:plot_order2}}
\end{center}     
\end{figure}

The integral of the probability density for $\alpha = 0.1$ defines a 90\% confidence range which is given by
\begin{eqnarray}
\int_{x_1}^{x_2} P(x,\mu,\sigma_\epsilon)\, dx & = & 1-\alpha, \label{eq:cl}
\end{eqnarray}
where, from now on, we assume that $x$ and $\mu$ are expressed in
units of $\sigma$. 
The condition on $x_1$ and $x_2$ in Eq.\,\ref{eq:cl} is satisfied for many pairs
$(x_1,x_2)$, and we select a particular pair using the  prescription from Feldman and Cousins in Ref.\,\cite{Feldman:1997qc}. 
The selection is
based on the ratio $R$ of the probability of measuring $x$, given the true mean value $\mu$ and scale uncertainty $\sigma_\epsilon$,
and the probability of obtaining $x$ given the best-fit physically allowable mean value 
$\mu$=max(0,$x$)\footnote{The value of $\mu$ corresponding to the maximum of the probability density actually has
a weak dependence on the nuisance parameters, but it is ignored here.}:
\begin{eqnarray}
R(x, \mu, \sigma_\epsilon) & = & 
{P(x| \,\mu, \sigma_\epsilon) \over P(x| \, \mu=x, \sigma_\epsilon) } \hspace{0.5cm} x \ge 0 \nonumber \\
& = & {P(x| \, \mu , \sigma_\epsilon) \over P(x| \,\mu=0, \sigma_\epsilon) } \hspace{0.5cm} x < 0. \label{eq:ordering}
\end{eqnarray}
The effect of using this ordering principle is to select central confidence intervals for large values of $x$ and
transitioning smoothly to upper confidence limits when $x$ is small or negative. The function $R(x)$ is 
a simple Gaussian for large values of $\mu$, but it becomes increasingly asymmetric as $\mu$ approaches zero.
This effect is illustrated for $\mu=0.2$ in Fig.\,\ref{fig:plot_order2} , where  $R(x)$ is plotted for four different values 
of the scale uncertainty $\sigma_\epsilon$. 

The pairs $(x_1,x_2)$ which satisfy both Eq.\,\ref{eq:cl} and Eq.\,\ref{eq:ordering} define the upper and lower limits
at the 90\% confidence level for a given true mean $\mu$  and scale uncertainty $\sigma_\epsilon$. For the
case considered by Feldman and Cousins in Ref.\cite{Feldman:1997qc}, which ignores scale uncertainties, 
the integrals in Eq.\,\ref{eq:cl} can be specified using the 
standard error function (and its inverse); the condition in Eq.\,\ref{eq:ordering} is satisfied
by requiring $\ln(R(x_1))$ = $\ln(R(x_2))$ \cite{FCclasnote}. This leads to
the following conditions:
\begin{eqnarray}
x_2 & = & 2\mu - x_1, \hspace{1.8cm} x_1 \ge 0   \nonumber \\
    & = & \mu + \sqrt{\mu^2 - 2 x_1 \mu}, \hspace{0.25cm} x_1 < 0. \label{eq:x2rx}
\end{eqnarray}
However, given the complexity of Eq.\,\ref{eq:fullprob} for non-zero $\sigma_\epsilon$, the integrals to 
determine the confidence intervals,
the computation of the ordering principle in Eq.\,\ref{eq:ordering} as well as the 
the simultaneous solution for pairs $(x_1,x_2)$ that satisfy both conditions have been evaluated numerically.

\section{Evaluation of confidence intervals}
\begin{figure}[tp]
\begin{center}
\vspace{-2cm}
\includegraphics[height=9cm,bb=20 20 502 522,clip=true]{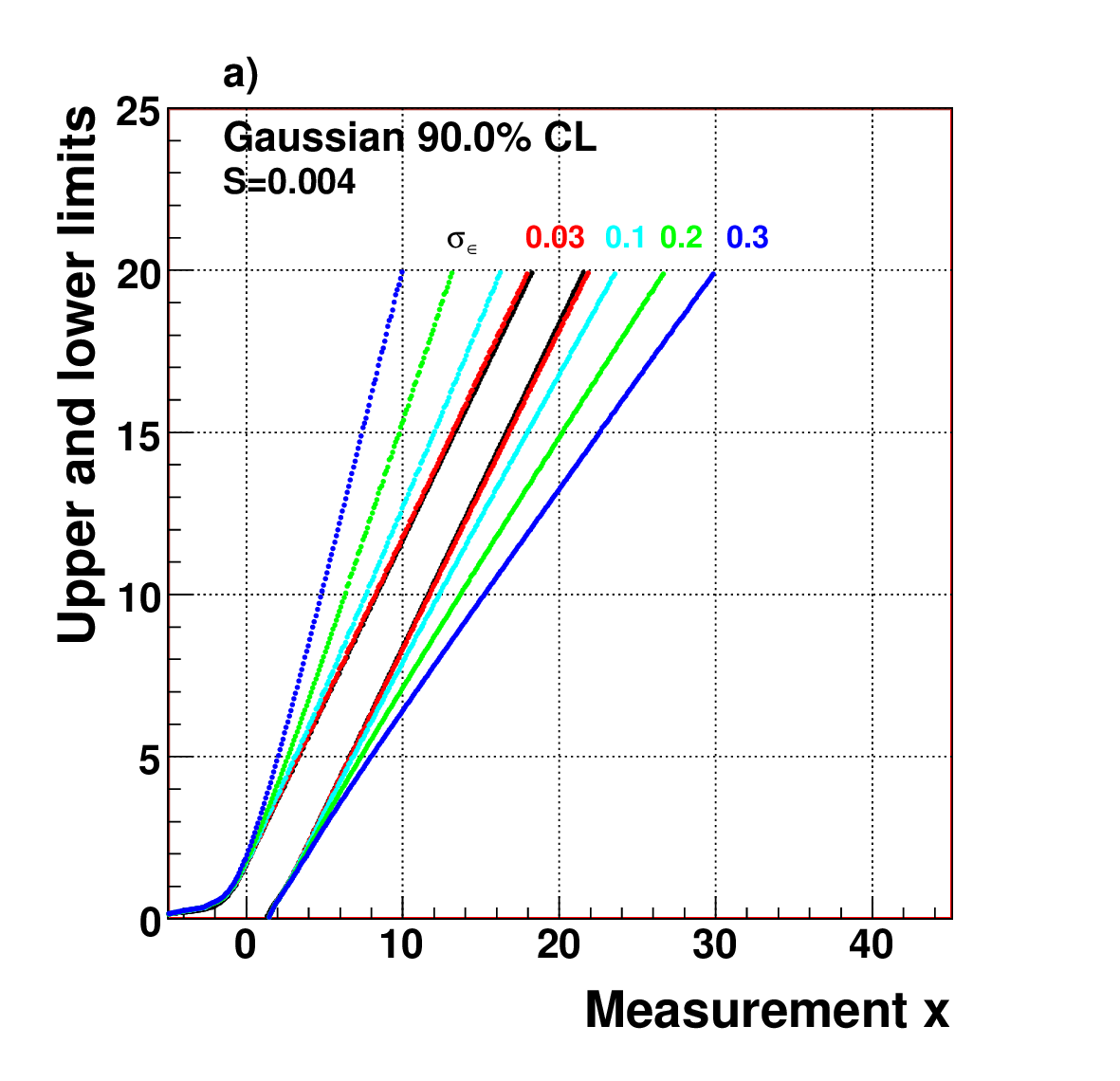}\\
\includegraphics[height=9cm,bb=20 20 502 522,clip=true]{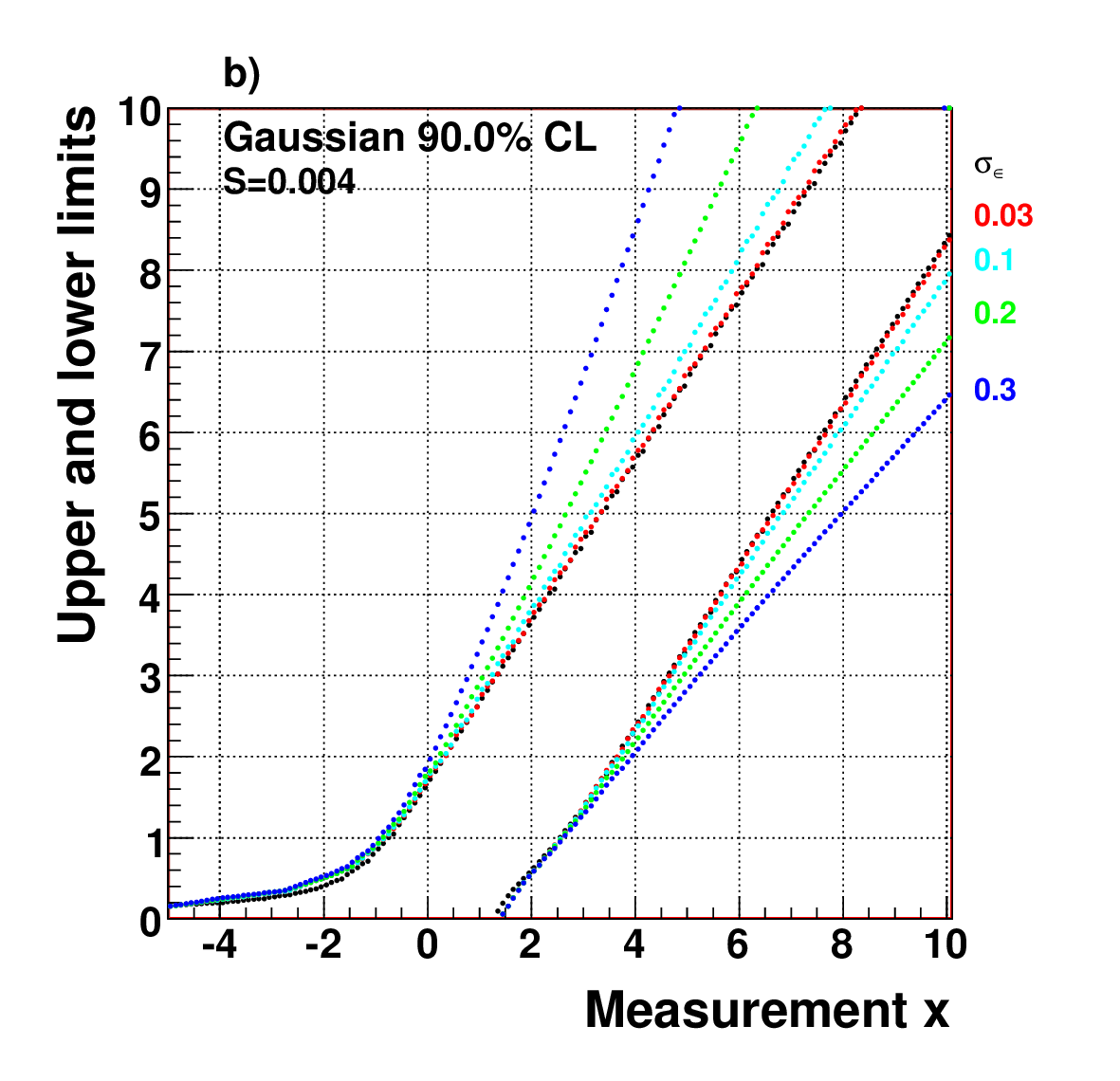} 
\caption{(Color online) The 90\% CL confidence intervals are plotted as a function of the measurement $x$.
Both axes are plotted in units of the standard deviation of the measurement $\sigma$.
The four limits correspond to relative scale uncertainties of $\sigma_\epsilon$ = 0.03, 0.1, 0.2 and 0.3,
respectively. The black curves show the standard Feldman-Cousins analysis, which ignores
systematic uncertainties altogether.
a) Limits plotted on a large scale, b) Limits plotted on a finer scale around x = 0. For negative
values of $x$, the curves converge to the standard Feldman-Cousins analysis.
\label{fig:ranges}}
\end{center}     
\end{figure}

The numerical computation of upper and lower limits was accomplished by using a series of ROOT scripts that
are described in Ref.\,\cite{FCsysclasnote}. The limits are computed for a fixed value of the nuisance scale 
parameter $\hat{S}$ and for fixed values of the assumed relative scale uncertainty $\sigma_\epsilon$.
As discussed previously, the calculations used the fixed value of $\hat{S}$=0.004 but are
 independent of the value of this parameter in the region of interest. The Neyman constructions of the confidence
intervals were computed at fixed values of the mean $\mu/\sigma$ in increments of 0.1. The range of 
the measurement $x/\sigma$ was from -5 to 95 with 1000 bins, corresponding to a granularity of 0.1. 
The limits are reported over the range of $x/ \sigma$ from -5 to 45. The additional range of the computation was necessary
to insure that the normalization integrals fully contained the probability distribution function. The resulting limits are shown 
in Fig.\,\ref{fig:ranges}. Fig.\,\ref{fig:ranges}\,a shows the full range for the calculation of the limits, and \ref{fig:ranges}\,b plots 
the limits in the region around $x=0$, where the prescription of the ordering principle of Feldman and Cousins 
deviates from the classical central intervals. Tables\,\ref{tab:limits1} and \ref{tab:limits2} tabulate the upper and lower
limits for $x$ between $-3$ and +3. Also plotted is the result of the limits from the standard Feldman and Cousins procedure,
which ignores scale uncertainties altogether. 
We see that our calculations converge to the standard prescription  for small values of the scale
uncertainty, as one would expect. However, this result
is not guaranteed and is an attractive feature of the present calculation.

\begin{table}[p]
\caption{Table of 90\% confidence intervals for several values of the relative scale uncertainty  $\sigma_\epsilon$.
The second column contains the results of our calculation of the standard Feldman-Cousins analysis,
which differs by a couple of percent from Table\,X in Ref. \cite{Feldman:1997qc}, presumably due to numerical approximations. 
All values are in units of the standard deviation $\sigma$.
\label{tab:limits1}
}
\begin{center}
\begin{tabular}{|l|c|c|c|c|c|c|}
\hline
\hline 
x	&	$\sigma_\epsilon$=0.0	&	$\sigma_\epsilon$=0.03	&	$\sigma_\epsilon$=0.1	&	$\sigma_\epsilon$=0.2	&	$\sigma_\epsilon$=0.3	\\ \hline \hline
-3.0	&	0.00--0.27	&	0.00--0.32 	&	0.00--0.32 	&	0.00--0.33 	&	0.00--0.33 	\\
-2.9	&	0.00--0.28	&	0.00--0.33 	&	0.00--0.33 	&	0.00--0.34 	&	0.00--0.34 	\\
-2.8	&	0.00--0.29	&	0.00--0.34 	&	0.00--0.34 	&	0.00--0.35 	&	0.00--0.35 	\\
-2.7	&	0.00--0.30	&	0.00--0.35 	&	0.00--0.35 	&	0.00--0.37 	&	0.00--0.38 	\\
-2.6	&	0.00--0.31	&	0.00--0.37 	&	0.00--0.37 	&	0.00--0.39 	&	0.00--0.40 	\\
-2.5	&	0.00--0.32	&	0.00--0.40 	&	0.00--0.40 	&	0.00--0.41 	&	0.00--0.43 	\\
-2.4	&	0.00--0.34	&	0.00--0.42 	&	0.00--0.43 	&	0.00--0.43 	&	0.00--0.45 	\\
-2.3	&	0.00--0.35	&	0.00--0.45 	&	0.00--0.45 	&	0.00--0.45 	&	0.00--0.47 	\\
-2.2	&	0.00--0.37	&	0.00--0.47 	&	0.00--0.47 	&	0.00--0.47 	&	0.00--0.49 	\\
-2.1	&	0.00--0.39	&	0.00--0.49 	&	0.00--0.49 	&	0.00--0.49 	&	0.00--0.52 	\\
-2.0	&	0.00--0.41	&	0.00--0.51 	&	0.00--0.51 	&	0.00--0.52 	&	0.00--0.54 	\\
-1.9	&	0.00--0.43	&	0.00--0.54 	&	0.00--0.54 	&	0.00--0.54 	&	0.00--0.56 	\\
-1.8	&	0.00--0.46	&	0.00--0.56 	&	0.00--0.56 	&	0.00--0.56 	&	0.00--0.59 	\\
-1.7	&	0.00--0.49	&	0.00--0.59 	&	0.00--0.59 	&	0.00--0.59 	&	0.00--0.62 	\\
-1.6	&	0.00--0.53	&	0.00--0.62 	&	0.00--0.62 	&	0.00--0.62 	&	0.00--0.65 	\\
-1.5	&	0.00--0.57	&	0.00--0.64 	&	0.00--0.64 	&	0.00--0.65 	&	0.00--0.70 	\\
-1.4	&	0.00--0.61	&	0.00--0.69 	&	0.00--0.69 	&	0.00--0.72 	&	0.00--0.76 	\\
-1.3	&	0.00--0.66	&	0.00--0.74 	&	0.00--0.75 	&	0.00--0.78 	&	0.00--0.80 	\\
-1.2	&	0.00--0.71	&	0.00--0.80 	&	0.00--0.81 	&	0.00--0.82 	&	0.00--0.84 	\\
-1.1	&	0.00--0.77	&	0.00--0.86 	&	0.00--0.87 	&	0.00--0.88 	&	0.00--0.91 	\\
-1.0	&	0.00--0.83	&	0.00--0.92 	&	0.00--0.92 	&	0.00--0.94 	&	0.00--0.99 	\\
-0.9	&	0.00--0.90	&	0.00--0.98 	&	0.00--0.98 	&	0.00--1.01 	&	0.00--1.07 	\\
-0.8	&	0.00--0.97	&	0.00--1.04 	&	0.00--1.04 	&	0.00--1.09 	&	0.00--1.13 	\\
-0.7	&	0.00--1.04	&	0.00--1.11 	&	0.00--1.13 	&	0.00--1.17 	&	0.00--1.22 	\\
-0.6	&	0.00--1.12	&	0.00--1.19 	&	0.00--1.20 	&	0.00--1.23 	&	0.00--1.31 	\\
-0.5	&	0.00--1.21	&	0.00--1.27 	&	0.00--1.28 	&	0.00--1.32 	&	0.00--1.41 	\\
-0.4	&	0.00--1.29	&	0.00--1.37 	&	0.00--1.38 	&	0.00--1.43 	&	0.00--1.53 	\\
-0.3	&	0.00--1.38	&	0.00--1.47 	&	0.00--1.48 	&	0.00--1.54 	&	0.00--1.65 	\\
-0.2	&	0.00--1.47	&	0.00--1.57 	&	0.00--1.59 	&	0.00--1.66 	&	0.00--1.77 	\\
-0.1	&	0.00--1.57	&	0.00--1.67 	&	0.00--1.69 	&	0.00--1.76 	&	0.00--1.84 	\\
~0.0	&	0.00--1.67	&	0.00--1.77 	&	0.00--1.79 	&	0.00--1.83 	&	0.00--1.97 	\\
\hline \hline
\end{tabular}
\end{center}
\end{table}

\begin{table}[p]
\caption{Table of 90\% confidence intervals for several values of the relative scale uncertainty  $\sigma_\epsilon$.
The second column contains the results of our calculation of the standard Feldman-Cousins analysis,
which differs by a couple of percent from Table\,X in Ref. \cite{Feldman:1997qc}, presumably due to numerical approximations. 
All values are in units of the standard deviation $\sigma$.
\label{tab:limits2}
}
\begin{center}
\begin{tabular}{|l|c|c|c|c|c|c|}
\hline
\hline 
x0	&	$\sigma_\epsilon$=0.0	&	$\sigma_\epsilon$=0.03	&	$\sigma_\epsilon$=0.1	&	$\sigma_\epsilon$=0.2	&	$\sigma_\epsilon$=0.3	\\ \hline \hline
0.1	&	0.00--1.77	&	0.00--1.86 	&	0.00--1.88 	&	0.00--1.92 	&	0.00--2.11 	\\
0.2	&	0.00--1.87	&	0.00--1.93 	&	0.00--1.95 	&	0.00--2.04 	&	0.00--2.24 	\\
0.3	&	0.00--1.97	&	0.00--2.02 	&	0.00--2.04 	&	0.00--2.15 	&	0.00--2.38 	\\
0.4	&	0.00--2.07	&	0.00--2.12 	&	0.00--2.15 	&	0.00--2.27 	&	0.00--2.52	\\
0.5	&	0.00--2.17	&	0.00--2.27 	&	0.00--2.31 	&	0.00--2.38 	&	0.00--2.66 	\\
0.6	&	0.00--2.27	&	0.00--2.36 	&	0.00--2.39 	&	0.00--2.50 	&	0.00--2.81 	\\
0.7	&	0.00--2.37	&	0.00--2.43 	&	0.00--2.45 	&	0.00--2.62 	&	0.00--2.96 	\\
0.8	&	0.00--2.47	&	0.00--2.52 	&	0.00--2.56 	&	0.00--2.74 	&	0.00--3.11 	\\
0.9	&	0.00--2.57	&	0.00--2.61 	&	0.00--2.71 	&	0.00--2.85 	&	0.00--3.26 	\\
1.0	&	0.00--2.67	&	0.00--2.77 	&	0.00--2.83 	&	0.00--2.97 	&	0.00--3.41 	\\
1.1	&	0.00--2.77	&	0.00--2.86 	&	0.00--2.90 	&	0.00--3.10 	&	0.00--3.57 	\\
1.2	&	0.00--2.87	&	0.00--2.93 	&	0.00--3.01 	&	0.00--3.22 	&	0.00--3.73 	\\
1.3	&	0.05--2.97	&	0.00--3.02 	&	0.00--3.14 	&	0.00--3.34 	&	0.00--3.88 	\\
1.4	&	0.14--3.07	&	0.06--3.18 	&	0.06--3.25 	&	0.06--3.46 	&	0.06--4.04 	\\
1.5	&	0.24--3.17	&	0.16--3.28 	&	0.16--3.35 	&	0.16--3.58 	&	0.16--4.21 	\\
1.6	&	0.33--3.27	&	0.26--3.37 	&	0.26--3.42 	&	0.26--3.71 	&	0.26--4.37 	\\
1.7	&	0.40--3.37	&	0.36--3.44 	&	0.36--3.57 	&	0.36--3.84 	&	0.35--4.54 	\\
1.8	&	0.46--3.47	&	0.45--3.52 	&	0.45--3.67 	&	0.44--3.96 	&	0.44--4.70 	\\
1.9	&	0.53--3.57	&	0.53--3.68 	&	0.53--3.78 	&	0.52--4.09 	&	0.52--4.86 	\\
2.0	&	0.60--3.67	&	0.60--3.78 	&	0.60--3.87 	&	0.59--4.21 	&	0.59--5.03 	\\
2.1	&	0.67--3.77	&	0.67--3.87 	&	0.67--3.94 	&	0.66--4.34 	&	0.66--5.21 	\\
2.2	&	0.74--3.87	&	0.75--3.94 	&	0.75--4.09 	&	0.75--4.47 	&	0.74--5.38 	\\
2.3	&	0.81--3.97	&	0.83--4.08 	&	0.83--4.19 	&	0.82--4.59 	&	0.80--5.55 	\\
2.4	&	0.89--4.07	&	0.91--4.18 	&	0.91--4.29 	&	0.89--4.72 	&	0.87--5.73 	\\
2.5	&	0.97--4.17	&	0.99--4.27 	&	0.98--4.36 	&	0.97--4.85 	&	0.95--5.90 	\\
2.6	&	1.04--4.27	&	1.06--4.33 	&	1.06--4.50 	&	1.05--4.98 	&	1.03--6.07 	\\
2.7	&	1.13--4.37	&	1.15--4.43 	&	1.15--4.61 	&	1.13--5.11 	&	1.10--6.26 	\\
2.8	&	1.21--4.47	&	1.23--4.58 	&	1.23--4.73 	&	1.20--5.24 	&	1.17--6.43 	\\
2.9	&	1.30--4.57	&	1.33--4.68 	&	1.32--4.81 	&	1.28--5.38 	&	1.25--6.61 	\\
3.0	&	1.39--4.67	&	1.43--4.77 	&	1.42--4.96 	&	1.37--5.51 	&	1.32--6.79 	\\
\hline \hline
\end{tabular}
\end{center}
\end{table}

\section{Coverage \label{sec:coverage}}
In a frequentist, or classical approach, 
coverage is one of the defining characteristics of a method for determining confidence limits. An ideal
method will have correct coverage. In the case of 90\% confidence limits, a method with correct coverage will return
limits which contain the true value of the parameter of interest in precisely 90\% of many
repeated trials or experiments. Neither overcoverage nor undercoverage are desirable for the method.
Overcoverage refers to intervals that contain the true value more often than 90\%, and undercoverage refers to
intervals that contain the true value less often. A general feeling among physicists is that ``erring on the
side of caution" (i.e. overcoverage) is acceptable, but in this case, the sensitivity of a particular experiment
is not being fully exploited. Undercoverage, on the other hand, can lead to incorrectly counting a fluctuation
as a true signal, which is clearly an undesirable outcome.

\begin{figure}[tp]
\begin{center}
\includegraphics[height=6.5cm,bb=80 20 550 500,clip=true]{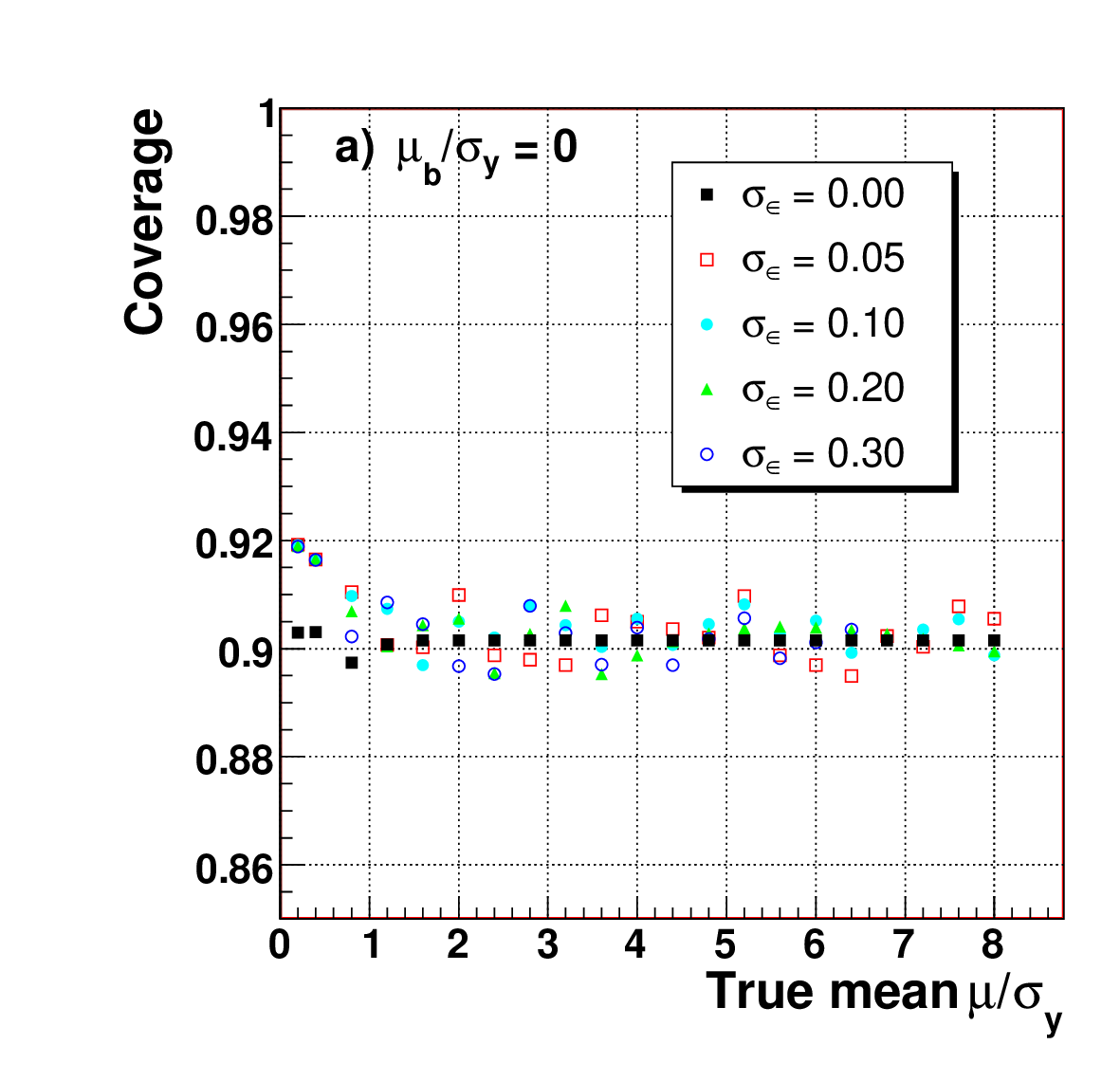}
\includegraphics[height=6.5cm,bb=80 20 550 500,clip=true]{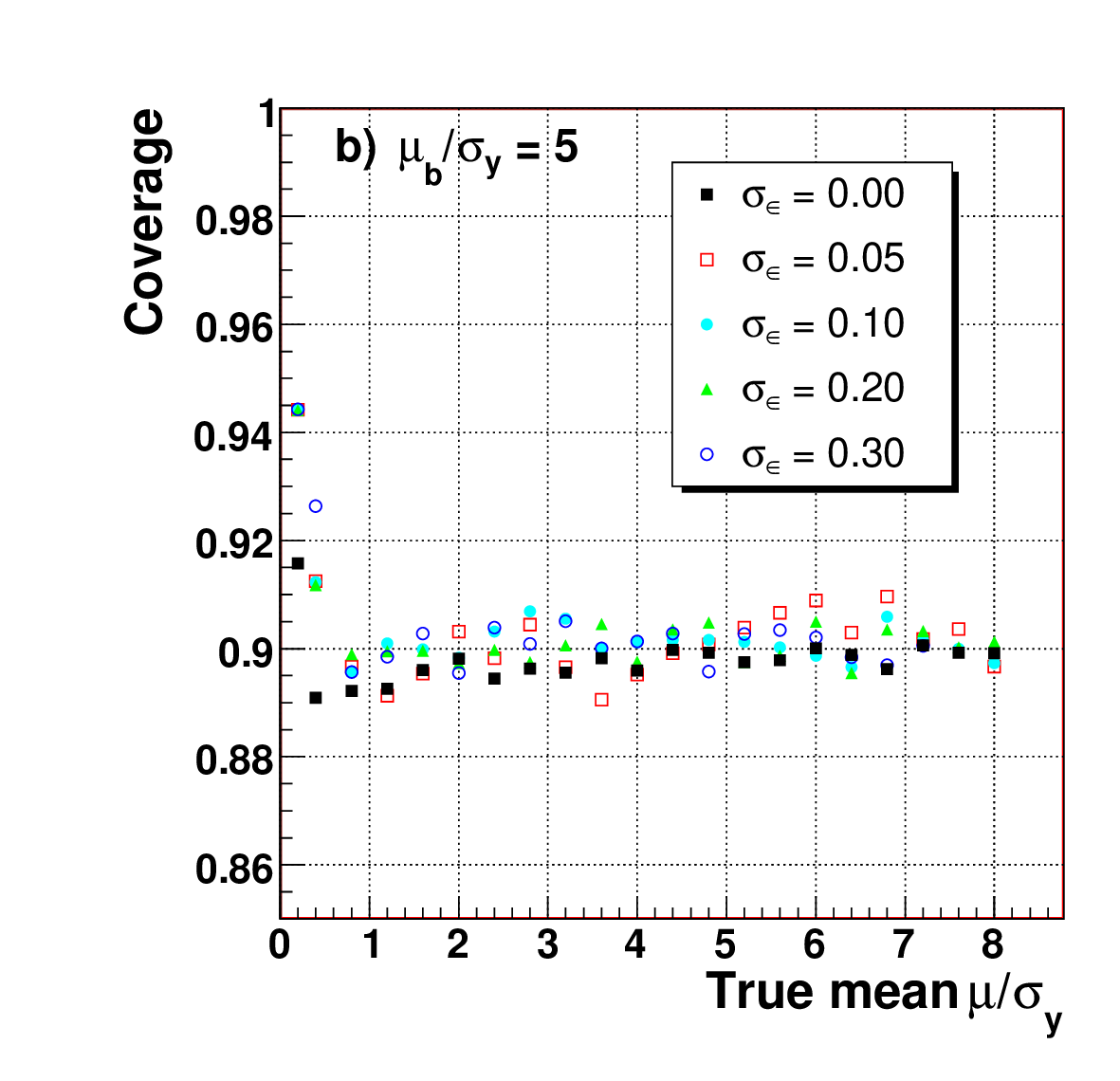}\\*[0.5cm]
\includegraphics[height=6.5cm,bb=80 20 550 500,clip=true]{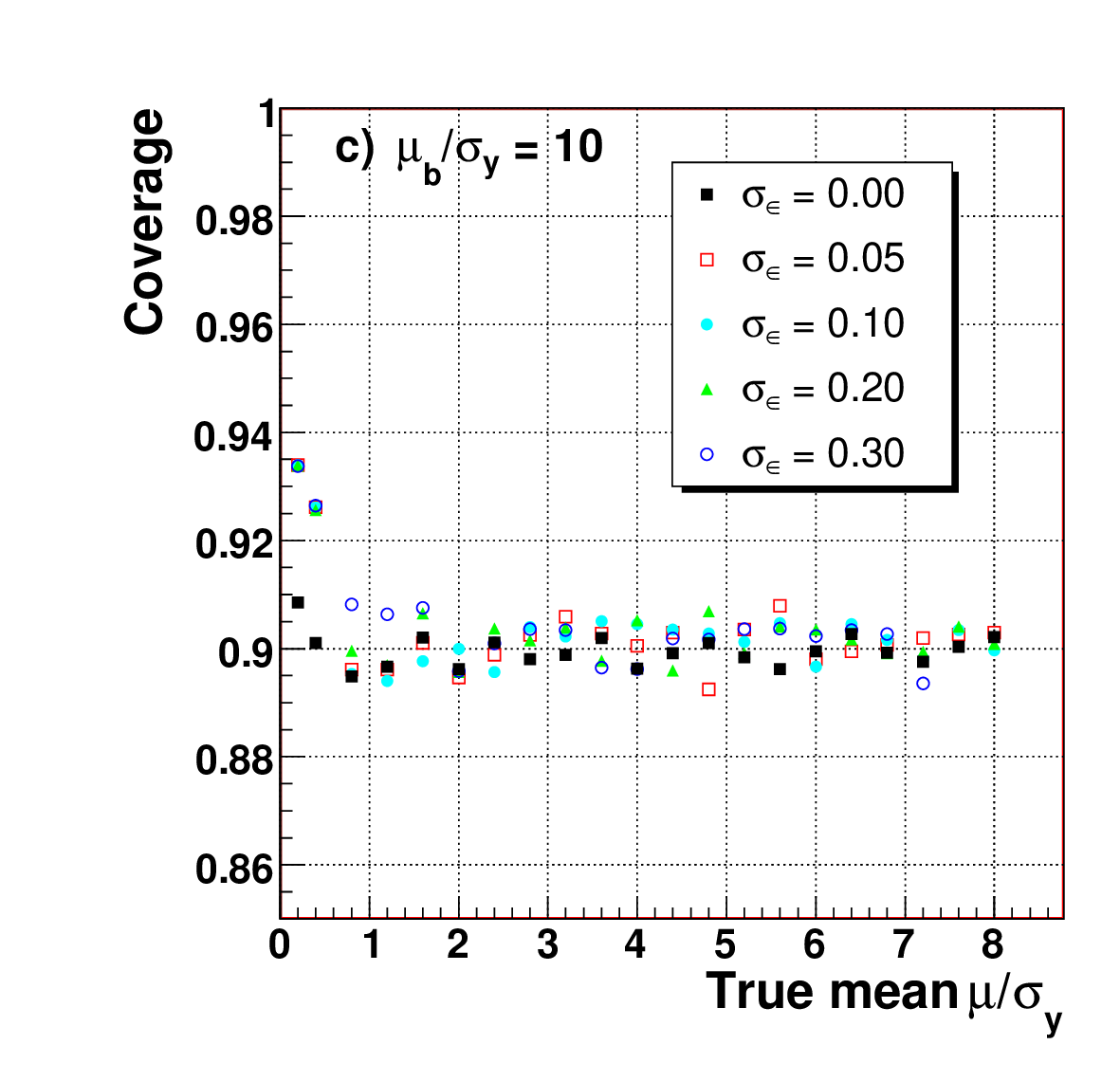}
\includegraphics[height=6.5cm,bb=80 20 550 500,clip=true]{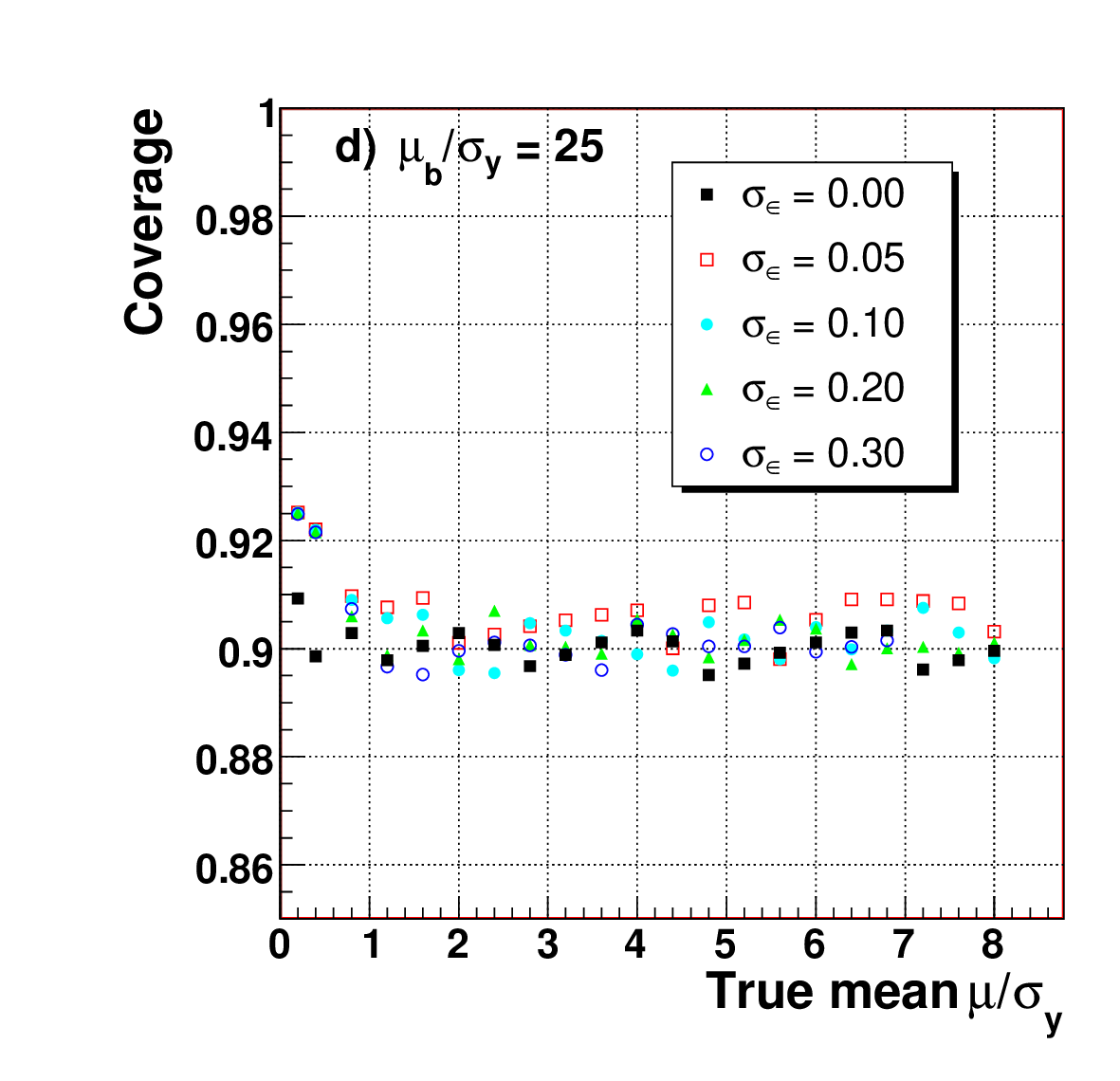}
\caption{(Color online) The computed coverage for 90\% confidence limits is shown for various values of the
scale uncertainty $\sigma_\epsilon$ at four values of the background level. The titles emphasize 
that variables are measured relative to the uncertainty (see text for details).
Most of the scatter about 0.90 is likely due to
numerical approximations in the determination of the confidence ranges. However, for 
small values of $\mu/\sigma_y$, the coverage
should approach the value given by the standard Feldman and Cousin result (shown
as solid squares) but in fact shows a systematic increase which is somewhat dependent on the
background level.
\label{fig:coverage}}
\end{center}     
\end{figure}

The coverage provided by the confidence intervals in our approach 
needs to be checked empirically. We note that
the standard Feldman-Cousins method for determining confidence intervals is known to have overcoverage
when applied to discrete distributions. This is due to the fact that the intervals are enlarged to
contain an integer number of counts, and this adjustment always increases coverage. 
For our case, where we consider only continuous variables $\mu$ and $x$, this is not a consideration.
But more to the point, the treatment of systematic uncertainties could change the coverage, and in
the Poisson case \cite{Tegenfeldt:2004dk,Heinrich:2004tj}, has also shown to produce overcoverage, typically
between 0.92 and 0.94. However, 
the coverage does depend on how one treats the nuisance variables. We show that as long as 
the  ``true'' efficiency $S$ and background level $\mu_b$ are treated as random variables, the 
generated intervals will have correct coverage over a large range of inputs.
In order to determine the coverage, we consider an ensemble of repeated Monte Carlo ``experiments"
performed for fixed values of the true parameter  $\mu$, the ``experimental'' unbiased estimator for the background level
$\mu_b$,  the measured detection efficiency $\hat{S}$
and the relative scale uncertainty $\sigma_\epsilon$.  For clarity in this section, we drop back to the notation from 
Section \ref{sec:prob}, where the uncertainties and scaling variables are shown explicitly.
The background level $\mu_b$ is assumed to be known
from other parasitic experiments or calibrations, or to have been calculated using theoretical input. Its uncertainty
must also be known or estimated. In our numerical examples, we take $\sigma_{\mu_b} = \sqrt{\mu_b}/5$.
Specifically, the following procedure was followed:
\begin{itemize}
\item The true efficiency $S$ was randomly selected from a Gaussian distribution $G(\hat{S},\hat{S}\,\sigma_\epsilon)$,
with mean $\hat{S}$ and a relative uncertainty $\sigma_\epsilon$.
\item The ``signal" $x$ was then selected from a Gaussian distribution $G(S \mu / \hat{S}, \sqrt{\mu})$. 
\item The true background level $b$ was selected from a Gaussian distribution $G(\mu_{b},\sigma_{\mu_{b}})$, 
\item The observed ``background'' $x_b$ was selected from a Gaussian distribution $G(b,\sqrt{\mu_{b}})$.
\item The reduced variable $y=(x+x_b-\mu_b)/\sigma_y$, where $\sigma_y=\sqrt{x + x_b + \sigma_{\mu_b}^2}$, 
was used to determined the
90\% CL for the specified value of $\sigma_\epsilon$,  returning the lower and upper limits $y^L$ and $y^U$.
\item If the true value  $\mu$ was contained in the confidence 
interval [$y^L\sigma_y,y^U\sigma_y]$, it was counted; 
otherwise, it was not.
\end{itemize}
At the end of a specified number of trials (one million), the coverage was calculated as the fraction of times the true
 value was contained within the confidence interval. The results are shown in 
 Fig.\,\ref{fig:coverage} at four values of the reduced background level
 $\mu_b/\sigma_y$ for various values of $\sigma_\epsilon$ .  The coverage for the case of $\mu_b/\sigma_y$=0
 was determined by generating the variable $y$ directly to avoid calculations where $\sigma_y=\sqrt{x}$ was very small.
 The figure shows that correct coverage is achieved within 1\% for $0.5 \leq \mu/\sigma_y \leq 10$. 
 For $\mu/\sigma_y$ less than 0.5, there is a systematic increase of the coverage that depends on the background level.
 The increase is due, in part,  to numerical
 approximations for large negative values of the measured quantity $x$, because for small $\mu$, the procedure itself should
 converge to the standard Feldman and Cousins result, which ignores systematic uncertainties and shows
 correct coverage.  Punzi  \cite{Punzi:2005yq} has studied the coverage for various ordering algorithms 
 and has shown that an ordering algorithm approximately based on the profile likelihoods can generate
 correct coverage for large values of the signal parameter practically independent of the scale uncertainty.
 However, for small values of the signal, which is governed by Poisson statistics, the coverage of this method 
 approaches unity.

\section{Example}
\begin{figure}[tp]
\begin{center}
\includegraphics[height=6.5cm,bb=20 20 502 502,clip=true]{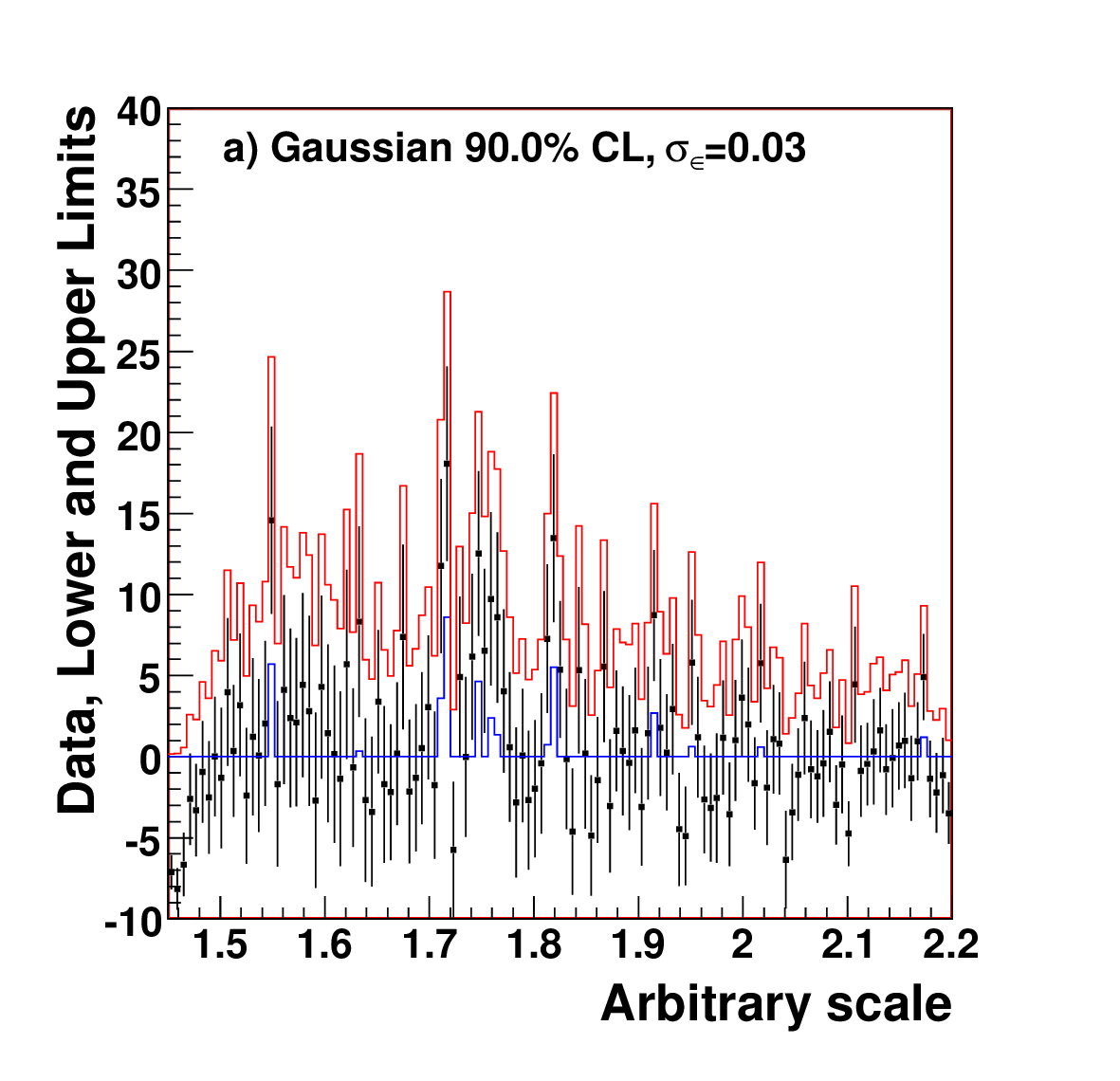}
\includegraphics[height=6.5cm,bb=20 20 502 502,clip=true]{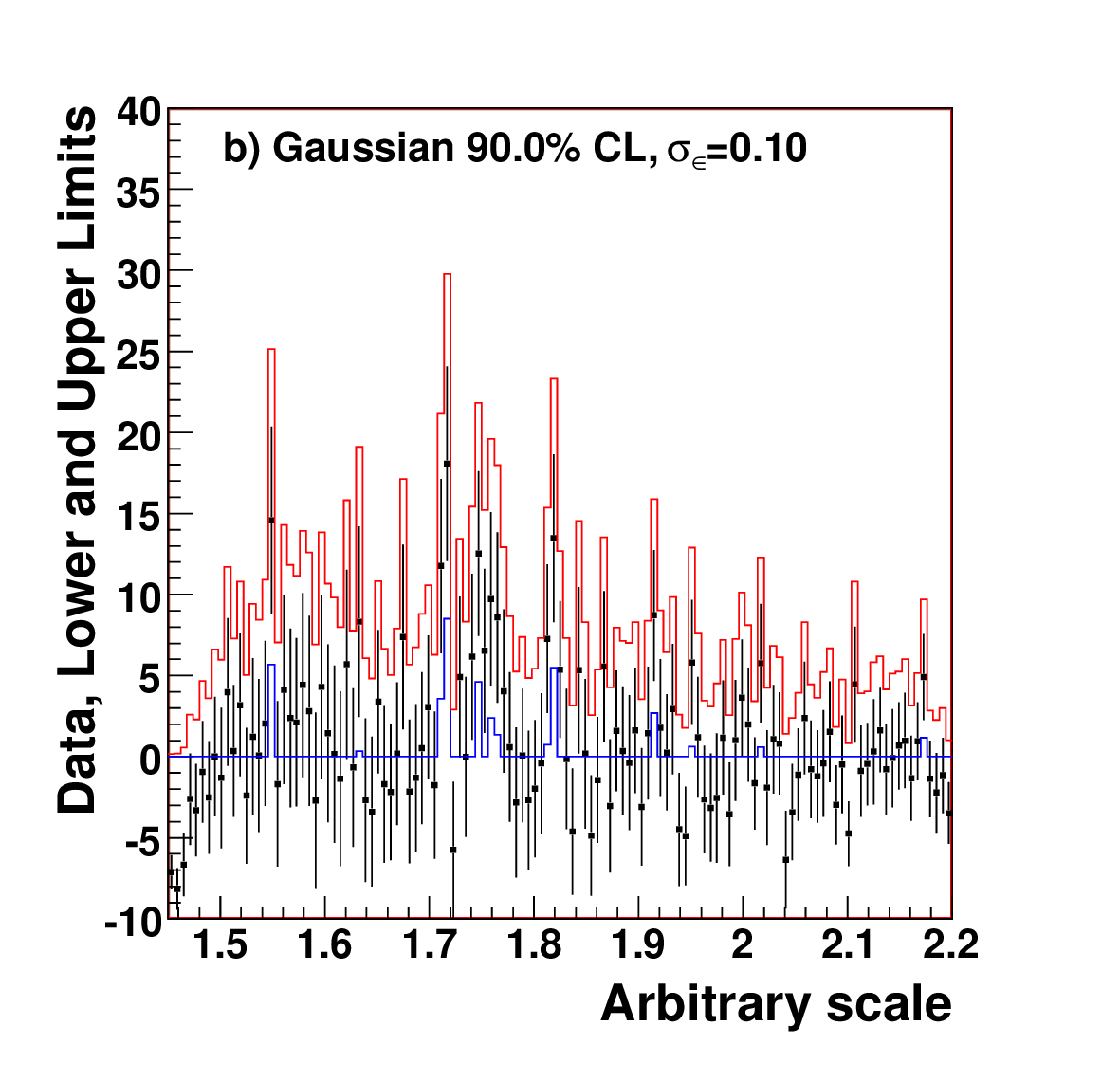} \\*[0.5cm]
\includegraphics[height=6.5cm,bb=20 20 502 502,clip=true]{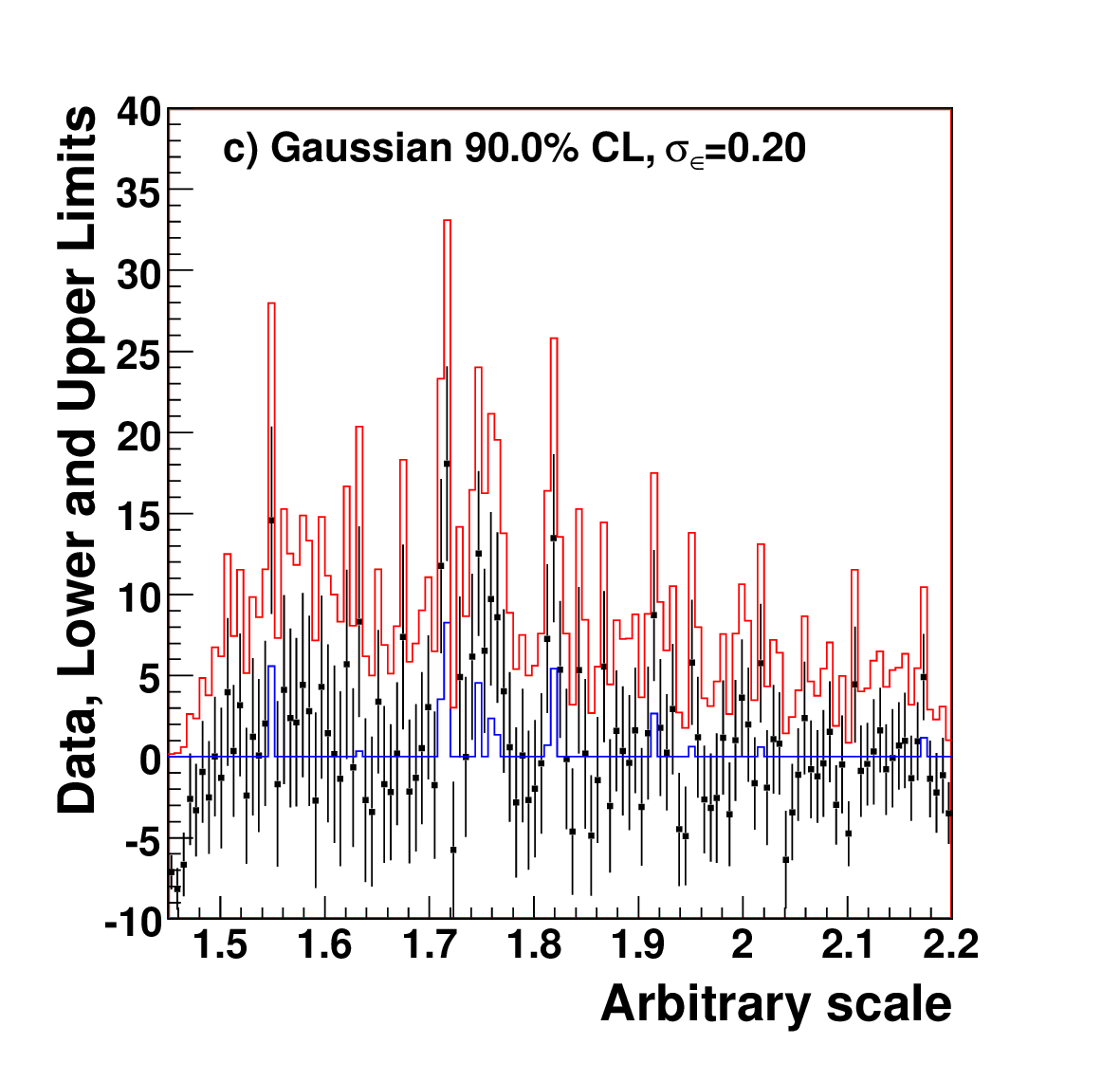}
\includegraphics[height=6.5cm,bb=20 20 502 502,clip=true]{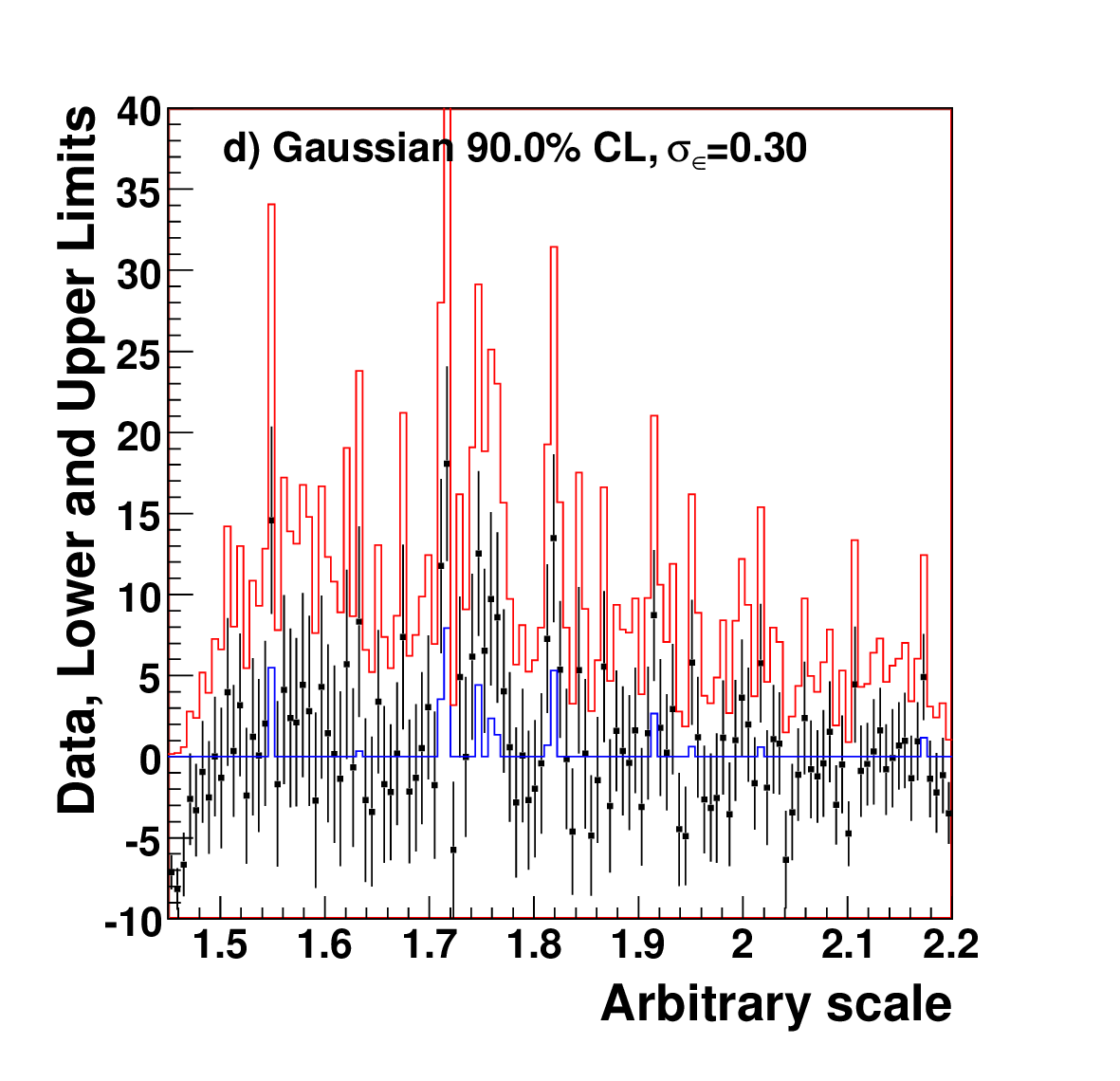}
\caption{(Color online) The upper and lower limits computed at the 90\% CL using four different assumptions for the
relative scale uncertainty $\sigma_\epsilon$ for the same sample of data a) $\sigma_\epsilon$=0.03,
b)  $\sigma_\epsilon$=0.10, c)   $\sigma_\epsilon$=0.20, and d) $\sigma_\epsilon$=0.30. The limits
in a) are very close to those obtained using the standard Feldman and Cousins analysis.
\label{fig:example_limits}}
\end{center}     
\end{figure}

By way of example, we take a histogram where each bin contains an estimate for the number of events
above background as a function of a variable that represents, for example, the kinematic coverage of the experiment. 
The number of signal events is negative when the estimated background exceeds the number of measured events,
and the errors on each point include the statistical uncertainties of the unsubtracted data as well as the uncertainty in 
the estimate of the mean background level (recall Eq.\,\ref{eq:gprob}). The same data with its errors is shown in Fig.\,\ref{fig:example_limits},
together with lower and upper limits determined with our method for four different assumptions for the relative
systematic uncertainty ($\sigma_\epsilon$=0.03, 0.10, 0.20 and 0.30). The plots provide a picture of how
the systematic uncertainties might change the limits that can be set using a given data set.
The limits on the first plot are very close to what
would be obtained with the standard Feldman and Cousins analysis. As the systematic error increases, so do the
upper limits.

\section{Dependence of limits on the scale uncertainties}
The application of our procedure for determining upper limits that we have described so far is quite
cumbersome. Many steps of computation are required to calculate the confidence
intervals, and even once determined, the limits are only valid for a fixed value of the relative
scale uncertainty. If limits are to be determined for a fixed value of $\sigma_\epsilon$, as in the 
previous example, the procedure is straightforward. However, it would not be unusual if
$\sigma_\epsilon$ depends, for example, on
the kinematics of the experiment, which would add considerable complexity to the
entire procedure.  Therefore, to streamline the calculations, we have investigated, and found,
a simple dependence of the limits on systematic uncertainties.

\begin{figure}[tp]
\begin{center}
\includegraphics[height=15cm,bb=23 24 490 505,clip=true]{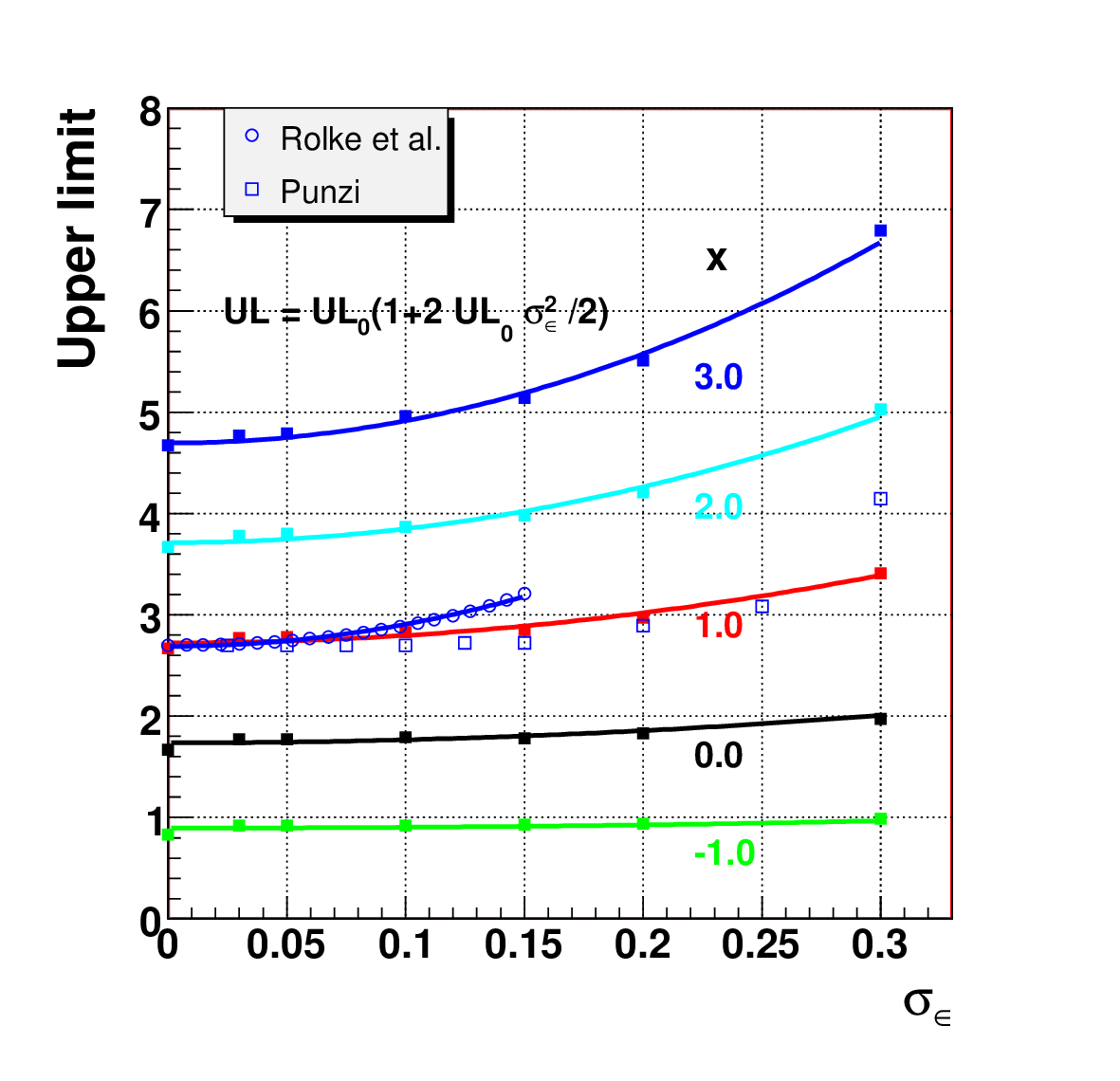}
\caption{(Color online) The 90\% upper limit is plotted as a function of the relative scale uncertainty $\sigma_\epsilon$
for various values of the measurement $x$. The standard Feldman and Cousins 
result, which ignores systematic uncertainties, is plotted for $\sigma_\epsilon$=0.
The dependence of the upper limit 
on the scale uncertainty computed  by Rolke et. al. \cite{Rolke:2004mj} (open circles)
and Punzi \cite{Punzi:2005yq} (open squares)
are also shown normalized to our upper limit for $x=1$. The data from Ref.\,\cite{Punzi:2005yq},
excluding the last point, are consistent with our analysis.
\label{fig:sigeps_dependence}}
\end{center}     
\end{figure}

The dependence of upper limits on the scale uncertainty is of interest in its own right and
was investigated in the original study by
Cousins and Highland in Ref.\,\cite{Cousins:1991qz}. In that low-counting application, they concluded
that the upper limit $U$ with a systematic uncertainty $\sigma_\epsilon$ can be calculated from
the limit $U_0$ with no systematic error using the following formula:
\begin{eqnarray}
U & = & U_{0} (1 + a\,U_{0}\; \sigma_\epsilon^2/2),
\label{eq:upar}
\end{eqnarray}
with $a$ =1.  Motivated by this study, we have fitted our results to Eq.\,\ref{eq:upar}. 
For the limited range of $x <$ 3, shown in Fig.\,\ref{fig:sigeps_dependence}, our results are
described quite well with the empirical value of $a$=2 if we substitute the standard Feldman and Cousins
result for $U_0$. We also note that this procedure has the desirable property of  continuity, 
that is it reduces to the Feldman and Cousins result for $\sigma_\epsilon$=0. 
For reference, we show the dependence obtained by Rolke and collaborators in 
Ref.\,\cite{Rolke:2004mj} and by Punzi \cite{Punzi:2005yq} normalized to our result at $x=1$, which roughly correspond to the
parameters for their study. The upper limits from Rolke show a dependence consistent with a stronger dependence
on the scale uncertainty with $a=6$, but the data from Punzi, except for their the last point, is in
very good agreement with our analysis.  It must be noted, however, that these comparisons can
only be qualitative, since their applications assume Poisson statistics.

\begin{figure}[tp]
\begin{center}
\includegraphics[height=15cm,bb=23 24 490 505,clip=true]{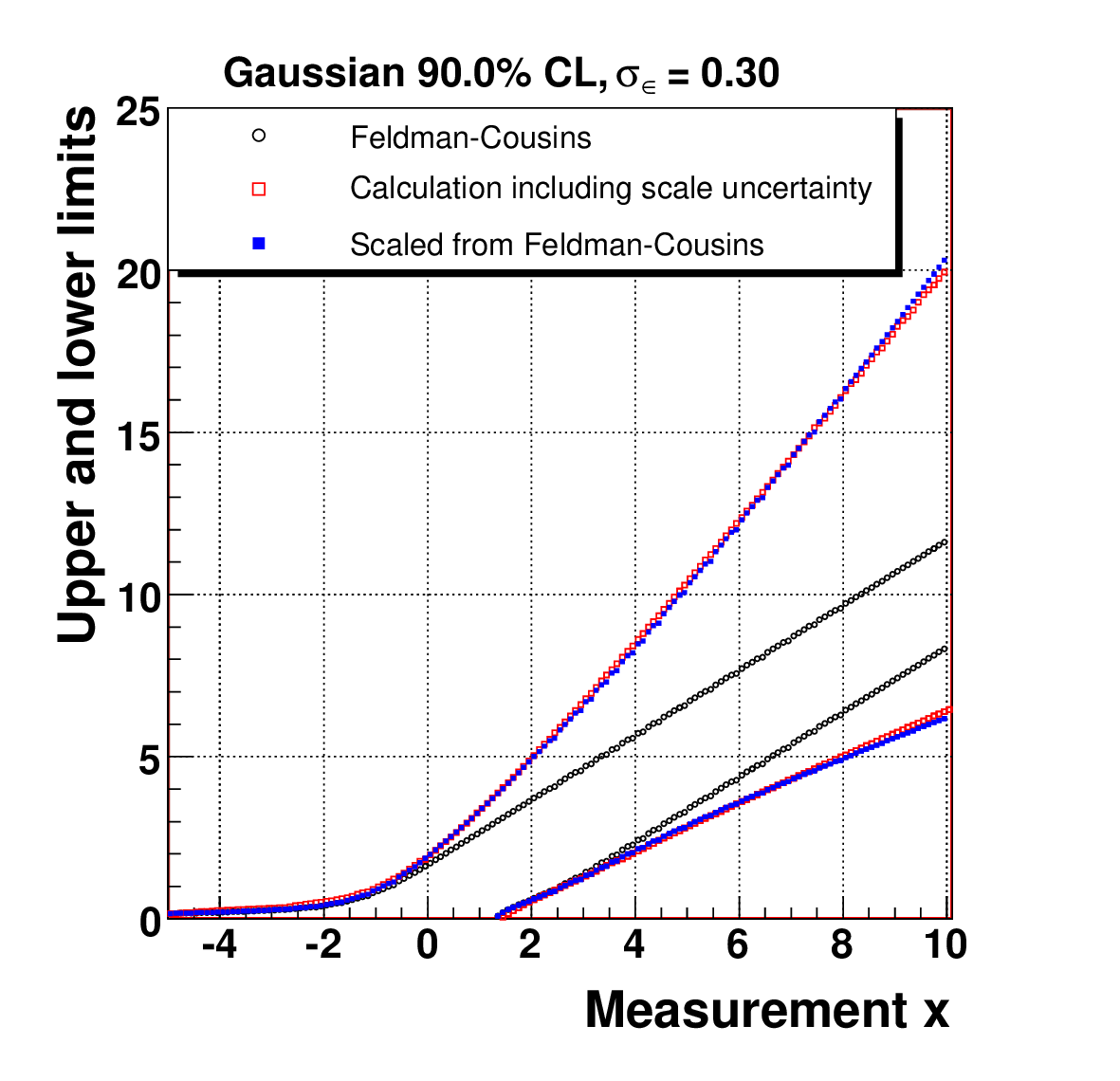}
\caption{(Color online) Three sets of confidence intervals are plotted. The central curves (open circles) 
reproduce the result of 
Feldman and Cousins from Ref.\,\cite{Feldman:1997qc}, which ignores systematic errors. 
The curves slightly inside the outer curves (open squares) are 
the calculated limits assuming a relative scale uncertainty of $\sigma_\epsilon$=0.3.
The outer curves (solid squares) are scaled from the Feldman-Cousins result using Eq.\,\ref{eq:upar}. 
The scaling provides a very good representation of the calculations, but deviates slightly
at $x \sim$ 10.
\label{fig:compare_limits_par}}
\end{center}     
\end{figure}

In order to cover a larger range of $x$ ($x<$10), it is necessary to extend the parameterization in Eq.\,\ref{eq:upar}
by allowing $a$ to be a function of $x$. Happily, the following common parameterization can be used for both
upper and lower limits:
\begin{eqnarray}
a (x) & = & A\, exp(-B x/2) \hspace{1cm} x > 2   \nonumber \\
a (x) & = & A\, exp(-B) \hspace{1.62cm} x < 2,
\end{eqnarray}
with $A_{U}$ = 2.26 and   $B_{U}$ = 0.092 assigned for upper limits, and  $A_{L}$ = $-$1.34 and  $B_{L}$ = 0.134 used
to determine the lower limits. Using this prescription, we compare our calculations for $\sigma_\epsilon$=0.3 
with the scaled Feldman-Cousins result from Ref.\,\cite{Feldman:1997qc} using Eq.\,\ref{eq:upar}. 
There is good agreement as shown in Fig.\,\ref{fig:compare_limits_par}. We also note that the
lower limit is relatively insensitive to the systematic error for $x$ less than 3. However, the 
upper limit begins deviating from the standard result for $x$ greater than 0.The deviations of the scaling
from the calculations are typically within the numerical accuracy of our calculations (0.1\,$\sigma$), but they
can be as large as 0.4\,$\sigma$ at some extremes.

\section{Summary and conclusions}
We have investigated how uncertainties in the estimation of the background levels and
detection efficiency affect the 90\% confidence intervals in the unified approach of
Feldman and Cousins from Ref.\,\cite{Feldman:1997qc}. The study has been conducted for
experiments where the number of detected events is
large and can be described by a Gaussian probability density function.
The construction of confidence intervals is classical or frequentist,  but
the systematic uncertainties are incorporated into the construction by
weighting the probability  distribution with an assumed density for the nuisance parameter
and then integrating over the nuisance parameter. This integration  adds a Bayesian flavor to
the approach as it  incorporates information about our belief about the detection 
efficiency and the background quantities.  We assume the nuisance parameters have 
Gaussian probability densities, and have studied the effect of relative scale uncertainties
in the range between 0~$\le \sigma_\epsilon \le$~0.3.

We find that the 90\% confidence intervals increase quadratically with the size of the
relative scale uncertainties (Eq.\,\ref{eq:upar}), as anticipated from previous work. 
The confidence intervals that incorporate scale uncertainties can be obtained
from the unified confidence intervals proposed by Feldman and Cousins in Ref.\,\cite{Feldman:1997qc}
with a simple scaling algorithm.  The new intervals have several attractive features.
For small values of the true mean $\mu$, the scale uncertainties do not affect the confidence 
intervals, as one might expect. By construction, the confidence intervals should have correct
coverage, and this has been verified by Monte Carlo calculations. Finally, in
the limit that $\sigma_\epsilon$ approaches 0, the intervals reproduce the intervals by Feldman and Cousins 
that ignore systematic uncertainties.

\section{Acknowledgments}
This work was supported by the U.S. Department of Energy contract DE-AC05-06OR23177, under which
Jefferson Science Associates, LLC operates the Thomas Jefferson National Accelerator Facility. We would like
to thank Hovanes Egiyan for useful discussion and valuable suggestions, and to Kandice Carter for 
helpful suggestions to improve the manuscript.

\bibliographystyle{unsrt}
                                                                                        
\bibliography{fc_nim_systematics}

\hyphenation{Post-Script Sprin-ger}
\begin{thebibliography}{10}

\bibitem{Heinrich:2007zz}
Joel Heinrich and L.~Lyons.
\newblock {Systematic errors}.
\newblock {\em Ann. Rev. Nucl. Part. Sci.}, 57:145--169, 2007.

\bibitem{Cousins:2005hc}
R.~D. Cousins.
\newblock {Treatment of nuisance parameters in high energy physics, and
  possible justifications and improvements in the statistics literature}.
\newblock Prepared for PHYSTATO5: Statistical Problems in Particle Physics,
  Astrophysics and Cosmology, Oxford, England, United Kingdom, 12-15 Sep 2005.

\bibitem{Rolke:2004mj}
Wolfgang~A. Rolke, Angel~M. Lopez, and Jan Conrad.
\newblock {Confidence Intervals with Frequentist Treatment of Statistical and
  Systematic Uncertainties}.
\newblock {\em Nucl. Instrum. Meth.}, A551:493--503, 2005.

\bibitem{Conrad:2002kn}
Jan Conrad, O.~Botner, A.~Hallgren, and Carlos Perez de~los Heros.
\newblock {Including systematic uncertainties in confidence interval
  construction for Poisson statistics}.
\newblock {\em Phys. Rev.}, D67:012002, 2003.

\bibitem{Tegenfeldt:2004dk}
Fredrik Tegenfeldt and Jan Conrad.
\newblock {On Bayesian treatment of systematic uncertainties in confidence
  interval calculations}.
\newblock {\em Nucl. Instrum. Meth.}, A539:407--413, 2005.

\bibitem{Feldman:1997qc}
Gary~J. Feldman and Robert~D. Cousins.
\newblock {A Unified approach to the classical statistical analysis of small
  signals}.
\newblock {\em Phys. Rev.}, D57:3873--3889, 1998.

\bibitem{Cousins:1991qz}
Robert~D. Cousins and Virgil~L. Highland.
\newblock {Incorporating systematic uncertainties into an upper limit}.
\newblock {\em Nucl. Instrum. Meth.}, A320:331--335, 1992.

\bibitem{convery}
M.R. Convery.
\newblock Incorporating multiplicative systematic errors in branching ratio
  limits.
\newblock SLAC-TN-03-001, February 2003.

\bibitem{stenson}
K.~Stenson.
\newblock A more exact solution for incorporating multiplicative systematic
  uncertainties in branching ratio limits.
\newblock [arXiv:physics/0605236], May 2006.

\bibitem{FCclasnote}
E.S. Smith.
\newblock Feldman-cousins method for gaussian statistics.
\newblock CLAS-NOTE 2007-019, September 2007.

\bibitem{FCsysclasnote}
E.S. Smith.
\newblock Multiplicative scale uncertainties in the unified approach of feldman
  and cousins for constructing confidence intervals.
\newblock CLAS-NOTE 2008-020, September 2008.

\bibitem{Heinrich:2004tj}
Joel Heinrich et~al.
\newblock {Interval estimation in the presence of nuisance parameters. 1.
  Bayesian approach}.
\newblock [arXiv:physics/0409129], 2004.

\bibitem{Punzi:2005yq}
Giovanni Punzi.
\newblock {Ordering algorithms and confidence intervals in the presence of
  nuisance parameters}.
\newblock [arXiv:physics/0511202], 2005.

\end{thebibliography}

 \newpage

 \newpage

 \newpage

 \newpage

 \newpage

 \newpage

 \newpage

\end{document}